\documentclass[12pt]{article} 

\usepackage{ae,lmodern}
\usepackage[applemac]{inputenc} 
\usepackage[T1]{fontenc}

\usepackage{epsfig}
\usepackage{graphicx}
\usepackage{comment}
\usepackage{latexsym}
\usepackage{hyperref}
\usepackage{amsmath}
\usepackage{mathrsfs}
\usepackage{color}
\usepackage{amsbsy}
\usepackage{amssymb}
\usepackage{amsthm}
\usepackage{amsfonts}
\usepackage{cite}
\usepackage{enumitem}

\newcommand{\be}[0]{\begin{equation}}
\newcommand{\ee}[0]{\end{equation}}

\renewcommand{\thefootnote}{\fnsymbol{footnote}}

\newcommand{\R}{\mathbb{R}}
\newcommand{\Z}{\mathbb{Z}}

\renewcommand{\O}{{\cal O}}

\newcommand{\sign}{{\rm sign}\,}
\newcommand{\cst}{{\rm cst.}}
\newcommand{\abs}{|}

\newcommand{\ie}{{\em i.e.} }
\newcommand{\eg}{{\em e.g.} }
\newcommand{\via}{{\it via} }
\newcommand{\apriori}{{\it a priori} }
\newcommand{\where}{\mbox{where}}

\newcommand{\when}{\mbox{when}}
\renewcommand{\and}{\mbox{and}}

\newcommand{\esp}{\phantom{\!\!\overset{\displaystyle \abs}{\abs}}}

\newcommand{\bm}{\boldmath} 

\newcommand{\F}{{\cal F}}
\newcommand{\N}{{\cal N}}

\newcommand{\K}{{\cal K}}
\renewcommand{\P}{{\cal P}}

\newcommand{\Ms}{M_{\rm s}}
\newcommand{\gs}{g_{\rm s}}
\newcommand{\nF}{n_{\rm F}}
\newcommand{\nB}{n_{\rm B}}
\newcommand{\Vone}{{\cal V}_{\mbox{\scriptsize 1-loop}}}

\catcode`\@=11
\def\marginnote#1{}
\newcount\hour
\newcount\minute
\newtoks\amorpm
\hour=\time\divide\hour by60 \minute=\time{\multiply\hour by60
\global\advance\minute by-\hour}
\edef\standardtime{{\ifnum\hour<12 \global\amorpm={am}%
        \else\global\amorpm={pm}\advance\hour by-12 \fi
        \ifnum\hour=0 \hour=12 \fi
        \number\hour:\ifnum\minute<10 0\fi\number\minute\the\amorpm}}
\edef\militarytime{\number\hour:\ifnum\minute<10 0\fi\number\minute}
\def\draftlabel#1{{\@bsphack\if@filesw {\let\thepage\relax
   \xdef\@gtempa{\write\@auxout{\string
      \newlabel{#1}{{\@currentlabel}{\thepage}}}}}\@gtempa
   \if@nobreak \ifvmode\nobreak\fi\fi\fi\@esphack}
        \gdef\@eqnlabel{#1}}
\def\@eqnlabel{}
\def\@vacuum{}
\def\draftmarginnote#1{\marginpar{\raggedright\scriptsize\tt#1}}
\def\draft{\oddsidemargin -.2truein
        \def\@oddfoot{\sl preliminary draft \hfil
        \rm\thepage\hfil\sl\today\quad\militarytime}
        \let\@evenfoot\@oddfoot \overfullrule 3pt
        \let\label=\draftlabel
        \let\marginnote=\draftmarginnote
   \def\@eqnnum{(\theequation)\rlap{\kern\marginparsep\tt\@eqnlabel}%
\global\let\@eqnlabel\@vacuum}  }
\def\thebibliography#1{
\vskip 0.5cm \centerline{\bf \Large References}
\list{
[\arabic{enumi}]}{\settowidth\labelwidth{[#1]}
\leftmargin\labelwidth
\advance\leftmargin\labelsep
\usecounter{enumi}}
\def\newblock{\hskip .11em plus .33em minus .07em}
\sloppy\clubpenalty4000\widowpenalty4000
\sfcode`\.=1000\relax}

\renewcommand{\theequation}{\arabic{section}.\arabic{equation}}
\renewcommand{\section}{\setcounter{equation}{0}\@startsection
{section}{1}{0mm}{-\baselineskip}{0.5\baselineskip} {\normalfont\Large\bfseries}}
\renewcommand{\subsection}{\@startsection
{subsection}{2}{0mm}{-\baselineskip}{0.5\baselineskip} {\normalfont\large\bfseries}}
\renewcommand{\subsubsection}{\@startsection
{subsubsection}{3}{0mm}{-\baselineskip}{0.5\baselineskip}
{\normalfont\normalsize\slshape}}

\begin{document}


\begin{titlepage}
\begin{flushright}
CPHT-RR087.112017, November  2017
\vspace{1.5cm}
\end{flushright}
\begin{centering}
{\bm\bf \Large Quantum no-scale regimes in string theory}

\vspace{5mm}

 {\bf Thibaut Coudarchet, Claude Fleming and Hervé Partouche
 \footnote{thibaut.coudarchet@ens-lyon.fr, claude.fleming@polytechnique.edu, herve.partouche@polytechnique.edu}}

 \vspace{1mm}

{Centre de Physique Théorique, Ecole Polytechnique\footnote{Unité mixte du CNRS et de l'Ecole Polytechnique, UMR 7644.}\\
F--91128 Palaiseau cedex, France}

\end{centering}
\vspace{0.1cm}
$~$\\
\centerline{\bf\Large Abstract}\\

\begin{quote}
{We show that in generic no-scale models in string theory, the flat, expanding cosmological evolutions found at the quantum level can be attracted to a ``quantum no-scale regime'', where the no-scale structure is restored asymptotically. In this regime, the quantum effective potential is dominated by the classical kinetic energies of the no-scale modulus and dilaton. We find that this natural preservation of the classical no-scale structure at the quantum level occurs when the initial conditions of the evolutions sit in a subcritical region of their space.  On the contrary, supercritical initial conditions yield solutions that have no analogue at the classical level. The associated intrinsically quantum universes are sentenced to collapse and their histories last finite cosmic times. Our analysis is done at 1-loop, in perturbative heterotic string compactified on tori, with spontaneous supersymmetry breaking implemented by a stringy version of the Scherk-Schwarz mechanism.}
\end{quote}




\end{titlepage}
\newpage
\setcounter{footnote}{0}
\renewcommand{\thefootnote}{\arabic{footnote}}
 \setlength{\baselineskip}{.7cm} \setlength{\parskip}{.2cm}

\setcounter{section}{0}


\section{Introduction}

Postulating the classical Lagrangian of the Standard Model in rigid Minkowski spacetime proved to be a very efficient starting point for computing quantum corrections. However, beyond this Standard Model, theories sometimes admit a gravitational origin. In particular, considering $\N=1$ supergravity models in dimension $d=4$, where local supersymmetry is spontaneously broken in flat space, and restricting the Lagrangians to the relevant operators gives  renormalizable classical field theories in rigid Minkowski spacetime, where supersymmetry is softly broken \cite{Barbieri:1982eh}. In that case, consistency of the picture should imply the possibility to commute the order of the above operations, namely first computing quantum corrections and then decoupling gravity.

To explore this alternative point of view in arbitrary dimension $d$, the classical supergravity theories may be viewed in the framework of no-scale models \cite{noscale} in string theory, for loop corrections to be unambiguously evaluated. By definition, the no-scale models are classical theories where local (extended) supersymmetry is (totally) spontaneously broken in flat space. In this context, the supersymmetry breaking scale is a scalar field which is a flat direction of a positive semi-definite classical potential. Therefore, if its vacuum expectation value is undetermined classically, a common wisdom is that this no-scale structure breaks down at the quantum level (see \eg \cite {Weinberg:1988cp}). 

One way to implement a spontaneous breaking of supersymmetry in string theory is \via coordinate-dependent compactification \cite{SSstring, Kounnas-Rostand}, a stringy version of the Scherk-Schwarz mechanism \cite{SS}. An effective potential is generated at 1-loop and is generically of order $\O(M^d)$, where $M$ is the supersymmetry breaking scale measured in Einstein frame. Assuming a mechanism responsible for the stabilization of $M$ (above 10 TeV for $d=4$) to exist, one then expects the quantum vacuum to be anti-de Sitter- or de Sitter-like, with no way to obtain a theory in rigid Minkowski space, once gravity is decoupled. Exceptions may however exist. In type II \cite{L=0} and open   \cite{1-L=0} string theory, the 1-loop effective potential $\Vone$ of some models vanishes at specific points in moduli space.  In heterotic string, the closest analogous models \cite{GrootNibbelink:2017luf} are characterized by equal numbers of {\em massless} bosons and fermions (observable and hidden sectors included), so that  $\Vone$ is exponentially suppressed when~$M_{(\sigma)}$, the supersymmetry breaking scale measured in $\sigma$-model frame, is below the string scale $\Ms$ \cite{Itoyama:1986ei, SNSM, FR}. These theories, sometimes referred as super no-scale models, can even be dual to the former, where $\Vone$ vanishes \cite{HarveyL=0}. However, all these particular type II, orientifold or heterotic models are expected to admit non-vanishing  or non-exponentially suppressed higher order loop corrections \cite{L2}, in which case they may lead to conclusions similar to those stated in the generic case. Moreover, the particular points in moduli space where $\Vone$ vanishes or is exponentially small are in most cases saddle points. As a consequence, moduli fields are destabilized and, even if their condensations induce a small mass scale $M_H<M$ such as the electroweak scale, the order of magnitude of $\langle \Vone\rangle$ ends up  being of order $\O(M^{d-2}M_H^2)$ \cite{SNSM}, which is still far too large to be compatible with flat space.

In the present work, we will not assume the existence of  a mechanism of stabilization of $M$ that would lead (artificially) to an extremely small cosmological constant. Instead, we take seriously the time-dependance of $M$ induced by the effective potential, in a cosmological setting. We show the existence of an attractor mechanism towards flat Friedmann-Lemaître-Robertson-Walker (FLRW) expanding universes, where the effective potential is dominated by the kinetic energies of $M$ and $\phi$, the dilaton field. Asymptotically, the cosmological evolution converges to that found in the classical limit, where the no-scale structure is exact. For this reason, we refer to this mechanism as an attraction to a ``quantum no-scale regime''. In these circumstances, flatness of the  universe is not destabilized by quantum corrections, which justifies that rigid Minkowski spacetime can be postulated in quantum field theory. We stress that even if the effective potential, which scales like $M^d$, is negligible from a cosmological point of view, the net value of the supersymmetry breaking scale $M$ remains a fundamental ingredient of the theory in rigid spacetime, since it determines the order of magnitude of all soft breaking terms. Note however that the analysis of the constraints raised by astrophysical observations about the constancy of couplings and masses, or the  validity of the equivalence principle, stand beyond the scope of the present work \cite{Damour:2002nv}.    

The above statements are shown in heterotic string compactified on a torus, with the total spontaneous breaking of supersymmetry implemented by a stringy Scherk-Schwarz mechanism \cite{SSstring, Kounnas-Rostand}. Actually, we  analyze a simplified model presented in Sect.~\ref{setup}, where only a small number of degrees of freedom are taken into account. To be specific, we consider in a perturbative regime the 1-loop effective action restricted to the scale factor $a$, as well as $M$ and $\phi$. In terms of canonical fields, the scalars can be described by  a ``no-scale modulus'' $\Phi$ with exponential potential, and a free scalar $\phi_\bot$. Notice that numerous works have already analyzed such systems, namely scalar fields with exponential potentials \cite{solV, DS}, sometimes as autonomous dynamical systems or by finding explicit solutions. Motivated by different goals, these studies often stress the onset of transient periods of accelerated cosmology. Such models have been realized by classical compactifications involving compact hyperbolic spaces, S-branes or non-trivial fluxes (field strengths) \cite{accel}. 

In the present paper, we find that the space of initial conditions of the equations of motion can be divided into two parts, and we present explicitly the resulting cosmologies in Sects.~\ref{sur}--\ref{part}.\footnote{A particular case in dimension 4 is already presented in Ref.~\cite{attra}, and describes the cosmological evolution of a universe at finite temperature $T$, when  $T\ll M$.} In the first region, which is referred as supercritical and exists only if $\Vone$ is negative, no classical limit exists. Thus, the universe is intrinsically quantum and its existence is found to be limited to a finite lapse of cosmic time. On the contrary, when the initial conditions sit in the so-called subcritical second region, the perturbative solutions can be seen as deformations of classical counterparts. It is in this case that attractions to quantum no-scale regimes take place. If, as mentioned before, the latter can correspond to flat  expanding evolutions, we also find that other quantum no-scale regimes exist, which describe a Big Bang (or Big Crunch by time reversal). Moreover, when $\Vone$ is positive, a short period of accelerated expansion can occur during the intermediate era that connects no-scale regimes of the two previous natures \cite{accel}. Whereas when $\Vone$ is negative, $M$ decreases as the universe expands and is thus forever climbing its potential \cite{DS}. Notice that this behaviour contradicts the naive expectation that $M$ should run away to infinity and lead to large, negative and \apriori non-negligible potential energy. We also point out that  the  above perturbative properties are expected to be robust when higher order loop corrections are taken into account.\footnote{This supposes the implementation of a regularization scheme to get rid off infrared divergences arising at genus $g$, when massless propagators and non-vanishing tadpoles at genus $g-1$ exist. This may be done by introducing a small mass gap by curving spacetime \cite{Kiritsis:1994ta}.} Finally, we summarize our results and outlooks in Sect.~\ref{cl}.


\section{The setup}
\label{setup}

In this section, we consider a simplified heterotic string no-scale model in dimension $d\ge 3$, in the sense that the dynamics of only a restricted number of light degrees of freedom is taken into account. Our goal is to derive the 1-loop low energy effective action and associated field equations of motion to be solved in the following sections. 

At the classical level, the background is compactified on $n\ge1$  circles of radii $R_i$, times a torus, 
\be
\label{fac}
\prod_{i=d}^{d+n-1}S^1(R_i)\times T^{10-d-n}\, .
\ee
The volume moduli of $T^{10-d-n}$ are supposed to be small enough for the lightest Kaluza-Klein (KK) mass scale $c\Ms$ associated with this torus to be very large, $c\lesssim 1$. On the contrary, the $n$ circles are supposed to be large, $R_i\gg 1$, and are used to implement a coordinate-dependent compactification responsible for the 
total spontaneous breaking of supersymmetry \cite{SSstring, Kounnas-Rostand}. In $\mbox{$\sigma$-model}$ frame, we define the resulting low supersymmetry breaking scale to be 
\be
M_{(\sigma)}\equiv{\Ms \over \bigg(\displaystyle \prod_{i=d}^{d+n-1} R_i\bigg)^{1\over n}}\ll c\Ms\, .
\ee 

At the quantum level, assuming a perturbative regime, an effective potential is generated at 1-loop \cite{Itoyama:1986ei, SNSM, Faraggi:2014eoa},
\begin{align}
\label{vi}
\Vone^{(\sigma)}&\equiv -{\Ms^d\over (2\pi)^d}\int_\F {d^2\tau\over 2\tau_2^2}\, Z\esp \nonumber\\
&= (\nF-\nB)\, v_{d,n} \, M_{(\sigma)}^d +\O\!\left((c\Ms M_{(\sigma)})^{d\over 2}\, e^{-c\Ms/M_{(\sigma)}}\right)\!,
\end{align}
where $Z$ is the genus-1 partition function and $\F$ is  the fundamental domain of $SL(2,\Z)$, parameterized by $\tau\equiv \tau_1+i\tau_2$. In the second expression, $\nF, \nB$ count the numbers of massless fermionic and bosonic degrees of freedom, while $v_{d,n}>0$ depends (when $n\geq 2$) on the $n-1$ complex structure moduli, $R_i/R_d$, $i=d+1, \dots, d+n-1$. The origins of the different contributions are the following: 

- The $\nB+\nF$ towers of pure KK modes associated with the massless states and arising from the $n$ large directions yield the term proportional to $M_{(\sigma)}^d$. 

- On the contrary, the pure KK towers based on the states at higher string oscillator level lead to the exponentially suppressed contribution.

- Finally, all states with non-trivial winding numbers along the $n$ large directions, as well as the unphysical \ie non-level matched states yield even more suppressed corrections, $\O\big(e^{-\Ms^2/M_{(\sigma)}^2}\big)$. 

\noindent Since we restrict in the present paper to the regime where $M_{(\sigma)}\ll c\Ms$, we will neglect from now on the exponentially suppressed terms. Splitting the dilaton field into a constant background and a fluctuation, $\phi_{\rm dil}\equiv \langle \phi_{\rm dil}\rangle+\phi$, the 1-loop low energy effective action restricted to the graviton, $\phi$ and the radii $R_i$'s takes the following form in Einstein frame\footnote{The 1-loop effective potential induces a backreaction implying a motion of the classical background. Adding the 1-loop correction to the kinetic terms is then unnecessary since it would introduce a correction to the cosmological evolution effectively at second order in string coupling $\gs$.}:  
\be
S={1\over \kappa^2}\int d^dx \, \sqrt{-g}\, \bigg[{{\cal R}\over 2}-{2\over d-2}(\partial\phi)^2-{1\over 2}\sum_{i=d}^{d+n-1}\left({\partial R_i\over R_i}\right)^2-\kappa^2\Vone\bigg].
\ee
In this expression,  ${\cal R}$ is the Ricci curvature, $\kappa^2=e^{2\langle \phi_{\rm dil}\rangle}/\Ms^{d-2}$ is the Einstein constant, and the potential  is dressed with the dilaton fluctuation, 
\be
\Vone\equiv e^{{2d\over d-2}\phi}\, \Vone^{(\sigma)}\simeq (\nF-\nB)\, v_{d,n}\,M^d\, ,
\ee 
where $M$ is the supersymmetry breaking scale measured in Einstein frame, 
\be
M\equiv e^{{2\over d-2}\phi}\, M_{(\sigma)}\, .
\ee
Note that the classical limit of the theory is recovered by taking $\kappa^2\to 0$.

In order to write the equations of motion, it may be convenient to perform field redefinitions.  The kinetic term of the scalar field $M$ being non-canonical, we define the so-called ``no-scale modulus'' $\Phi$ as
\be
M\equiv e^{\alpha\Phi}\Ms \qquad i.e.  \qquad \alpha \Phi= {2\over d-2}\,  \phi-{1\over n}\sum_{i=d}^{d+n-1}\ln R_i\, , 
\ee
where $\alpha$ is an appropriate normalization factor, 
\be
\label{al}
\alpha=\sqrt{{1\over d-2}+{1\over n}}\, . 
\ee
Moreover, the effective potential being by construction independent on the orthogonal combination
\be
\phi_\bot={1\over \sqrt{d-2+n}} \Big(2 \phi+\sum_{i=d}^{d+n-1}\ln R_i\Big),
\label{do}
\ee
the latter is a canonical free field.
By also redefining the complex structure deformations as
\begin{align}
\varphi_k={1\over \sqrt{k(k+1)}}\Big(k\ln R_{d+k}- \sum_{i=d}^{d+k-1}\ln R_i\Big) , \qquad k=1,\dots, n-1\, , 
\end{align}
the action takes the final form 
\be
S={1\over \kappa^2}\int d^dx \, \sqrt{-g}\, \bigg[{{\cal R}\over 2}-{1\over2}(\partial\Phi)^2-{1\over 2}(\partial\phi_\bot)^2-{1\over 2}\sum_{k=1}^{n-1}(\partial\varphi_k)^2-\kappa^2\Vone\bigg],
\ee
where the 1-loop effective potential is
\be
\Vone=  (\nF-\nB)\, v_{d,n}(\varphi_1,\dots,\varphi_{n-1})\,e^{d\alpha\Phi}\, \Ms^d \, .
\ee

To keep the toy model as simple as possible, we treat the complex structures as constants, $\varphi_k\equiv\cst$, $k=1,\dots, n-1$, and  ignore as well the remaining internal moduli (other than the volume $\prod_{i=d}^{d+n-1}R_i$ appearing in the definitions of $\Phi$ and $\phi_\bot$).  Looking for homogeneous and isotropic cosmological evolutions in flat space, we consider the metric and scalar field anzats\footnote{When $d=4$, we also take the axion field dual to the spacetime antisymmetric tensor to be constant.}  
\be
ds^2=-N(x^0)^2(dx^0)^2+a(x^0)^2\Big((dx^1)^2+\cdots+(dx^{d-1})^2\Big)\, , \quad \Phi(x^0)\, , \quad \phi_\bot(x^0)\, .
\ee
The equations of motion for the lapse function $N$, scale factor $a$, no-scale modulus $\Phi$ and $\phi_\bot$ take the following forms, in the gauge $N\equiv 1$ which defines cosmic time $x^0\equiv t$,
\begin{align}
{1\over 2}\, (d-1)(d-2)\, H^2&=\K + \kappa^2\Vone\, ,  \qquad  \K={1\over2}\, \dot\Phi^2+{1\over 2}\, \dot\phi_\bot^2\, , 
\label{fri}\\
(d-2)\, \dot H+{1\over 2}\, (d-1)(d-2)\, H^2&=-\K+\kappa^2\Vone\, , \label{e2}\\
\ddot\Phi+(d-1)\, H\, \dot\Phi &= -d\alpha  \kappa^2\Vone\, , \label{e3}\\
\ddot\phi_\bot+(d-1)\, H\, \dot \phi_\bot&=0\, , \label{e4}
\end{align}
where $H\equiv \dot a/a$. In order to solve the above differential system, we consider a linear combination of the three first equations that eliminates both $\K$ and $\Vone$,
\be
\label{to}
\Big(\alpha \dot\Phi+{\alpha^2\over 2}\, d(d-2)\,H\Big)^{\displaystyle\cdot} +(d-1)\, H\Big(\alpha \dot\Phi+{\alpha^2\over 2}\, d(d-2)\, H\Big)=0\, , 
\ee
which is a free field equation identical to that of $\phi_\bot$. Integrating, we have 
\be 
\dot\phi_\bot=\sqrt{2}\, {c_\bot\over a^{d-1}}\, , \qquad \alpha\dot\Phi+{\alpha^2\over 2}\, d(d-2)\,  H= {c_\Phi\over a^{d-1}}\, ,
\label{cc}
\ee
where $c_\bot, c_\Phi$ are arbitrary constants. Note that under time-reversal, the constants  $c_\Phi,c_\bot$ change to $-c_\Phi,-c_\bot$.  To proceed, it is useful to eliminate the effective potential between Eqs~(\ref{fri}) and~(\ref{e2}),
\be
\label{dotH}
{1\over 2}\, (d-2)\, \dot H=-\K\, .
\ee
The above equation is however a consequence of the others, as can be shown by  taking the time derivative of Eq.~(\ref{fri}) and using the scalar equations~(\ref{e3}),~(\ref{e4}).  Therefore, we can solve it and  insert the solution in Friedmann equation~(\ref{fri}) in order to find, when $\nF-\nB\neq 0$, the fully integrated expression of the no-scale modulus or $M$. 

To  reach this goal, we first use Eqs~(\ref{cc}) to express the scalar kinetic energy $\K$ as a function of the scale factor and $H$, so that Eq.~(\ref{dotH}) becomes a second order differential equation in $a$  only. Second, when $c_\Phi\neq 0$, we introduce a new (dimensionless)  time variable~$\tau$, in terms of which this equation becomes
\be
\label{eq}
\tau d\tau=-A \, \P(\tau)\, {da\over a}\, , \qquad \P(\tau)=\tau^2-2\tau+\omega\Big[1+2\alpha^2\Big({c_\bot\over c_\Phi}\Big)^2\Big]\, ,  
\ee
where we have defined 
\be
\tau\equiv {2A\over dc_\Phi}\, \dot a \, a^{d-2}\, , \qquad A={\omega\over 4}\, d^2(d-2)\alpha^2 \, , \qquad \omega=1-{4(d-1)\over d^2(d-2)\alpha^2}\, .
\label{tau}
\ee
Note that using the definition of $\alpha$ in Eq.~(\ref{al}), we have $0<\omega<1$, for arbitrary $d\ge 3$ and $n\ge 1$. Finally, using again Eqs~(\ref{cc}), Friedmann equation~(\ref{fri}) takes an algebraic form, once expressed in terms of time $\tau$,
\be
(\nF-\nB)\, v_{d,n}\,\kappa^2M^d=-{c_\Phi^2\over 2\alpha^2\omega}\, {\P(\tau)\over a^{2(d-1)}}\,.
\label{fri2}
\ee

We will see that the forms of the solutions for the scale factor $a$ and the supersymmetry breaking scale $M$ depend drastically on the number of real roots allowed by the quadratic polynomial $\P(\tau)$. Moreover, in order to find the restrictions for  string perturbation theory to be valid, we will need to display the dilaton field evolution. Using the definitions of the scalars $\Phi$ and $\phi_\bot$,  we have
\be
\label{dil}
e^{2d\alpha^2\phi}=e^{d\alpha\Phi}\, e^{ {d\over n}\sqrt{d-2+n}\,\phi_\bot}\, , 
\ee
where $\phi_\bot$ is determined by its cosmic time derivative, or
\be
\label{bot}
{d\phi_\bot\over d\tau}\, \P(\tau)=-{2\sqrt{2}\over d}\, {c_\bot\over c_\Phi}\, .
\ee
To derive the above relation, we have used the definition of $\tau$  and    Eq.~(\ref{eq}) to relate the time variables $t$ and $\tau$,
\be
{d\tau\over dt} = -{d\over 2}\, {c_\Phi\over a^{d-1}} \, \P(\tau)\, .
\ee

In the following sections, we describe the cosmological evolution obtained for arbitrary $c_\bot/c_\Phi$, which admits a critical value 
\be
\gamma_c=\sqrt{1-\omega\over 2\alpha^2\omega}\, , 
\label{gam}
\ee
corresponding to a null discriminant for $\P(\tau)$. 


\section{Supercritical case}
\label{sur}

When $c_\bot$ and $c_\Phi\neq 0$ satisfy the supercritical condition
\be
 \left\abs{c_\bot\over c_\Phi}\right\abs>\gamma_c\, , 
\ee
$\P(\tau)$ has no real root.  Due to Friedmann equation~(\ref{fri2}),  the no-scale model must satisfy $\nF-\nB<0$. Moreover, the classical limit $\kappa^2\to 0$ is not allowed (!) This very fact  means that in the case under consideration, the cosmological evolution of the universe is intrinsically driven by quantum effects. In particular, the time-trajectory cannot allow any regime where the 1-loop effective potential may be neglected. 

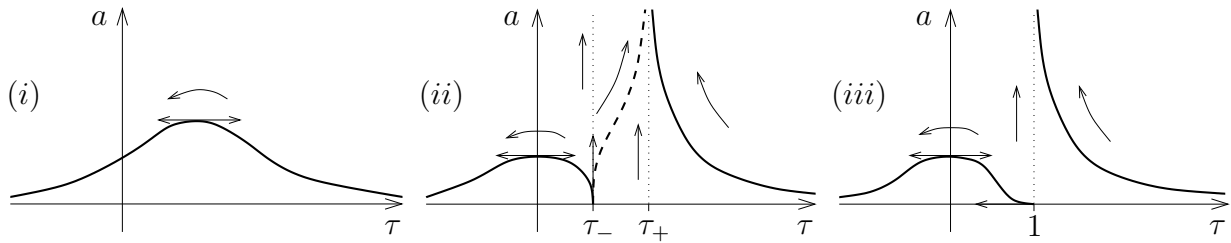
\begin{figure}[h!]
\vspace{.6cm}
\begin{center}
\includegraphics[height=3.0cm]{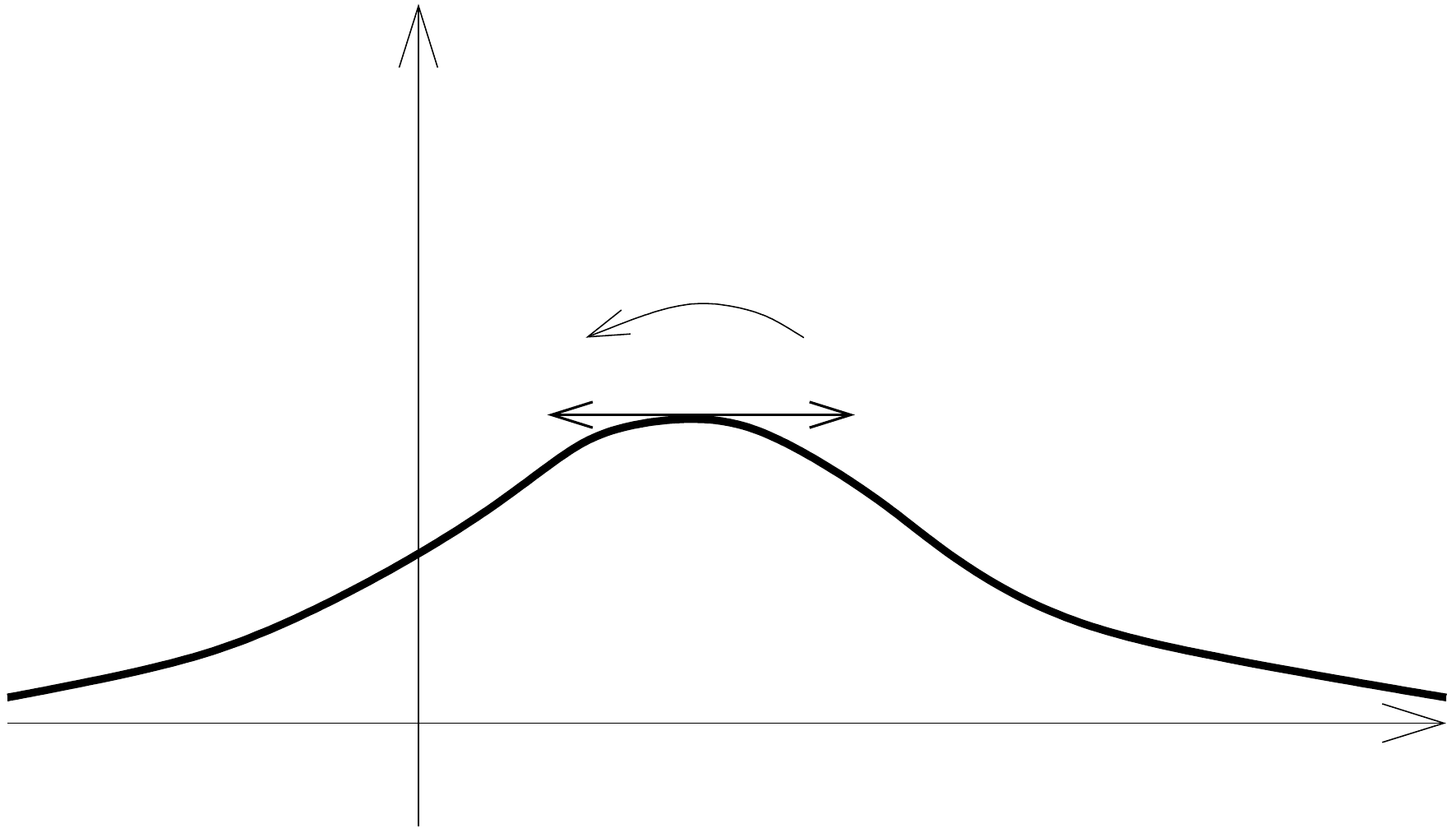}\;
\includegraphics[height=3.0cm]{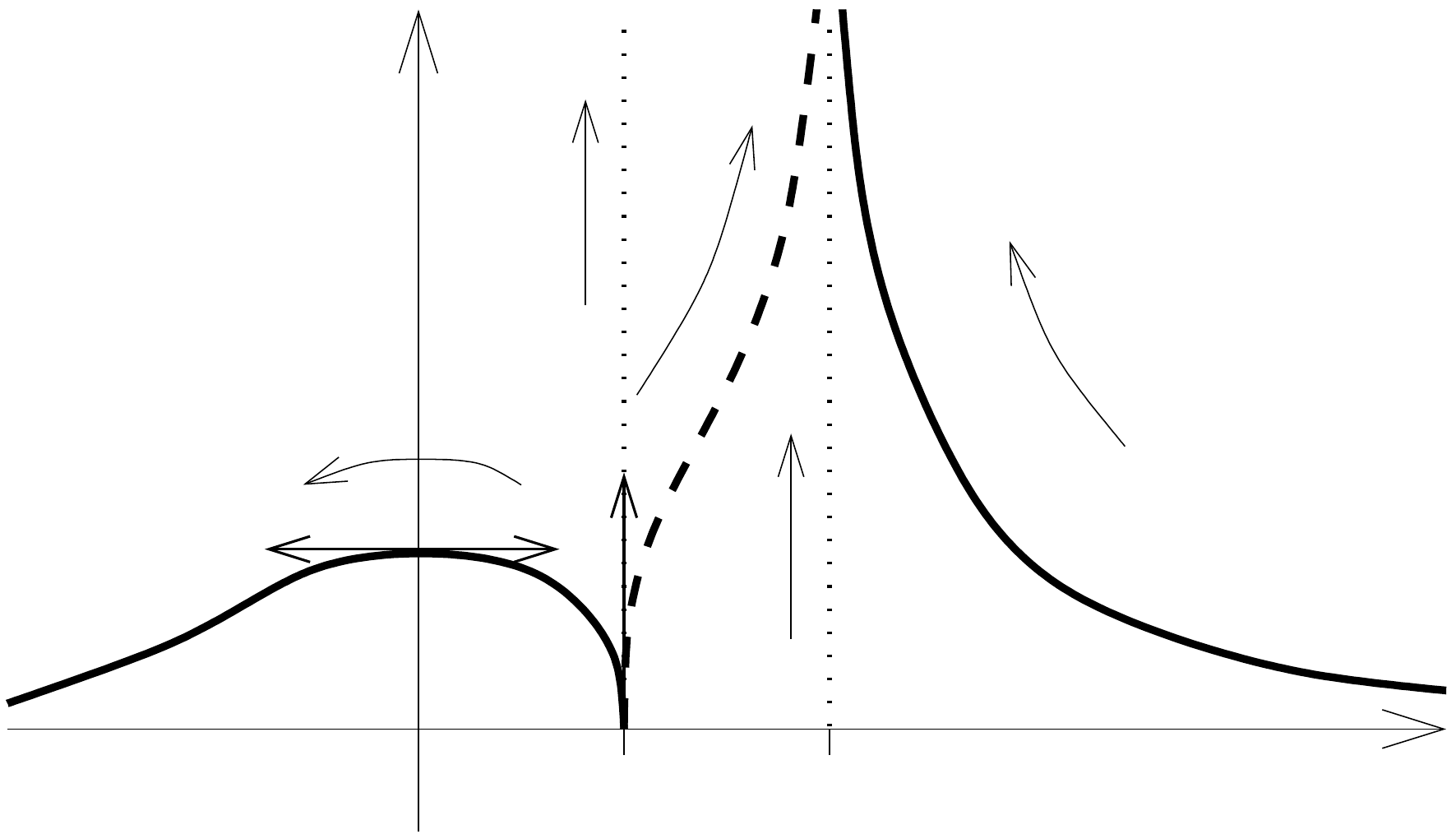}\;
\includegraphics[height=3.0cm]{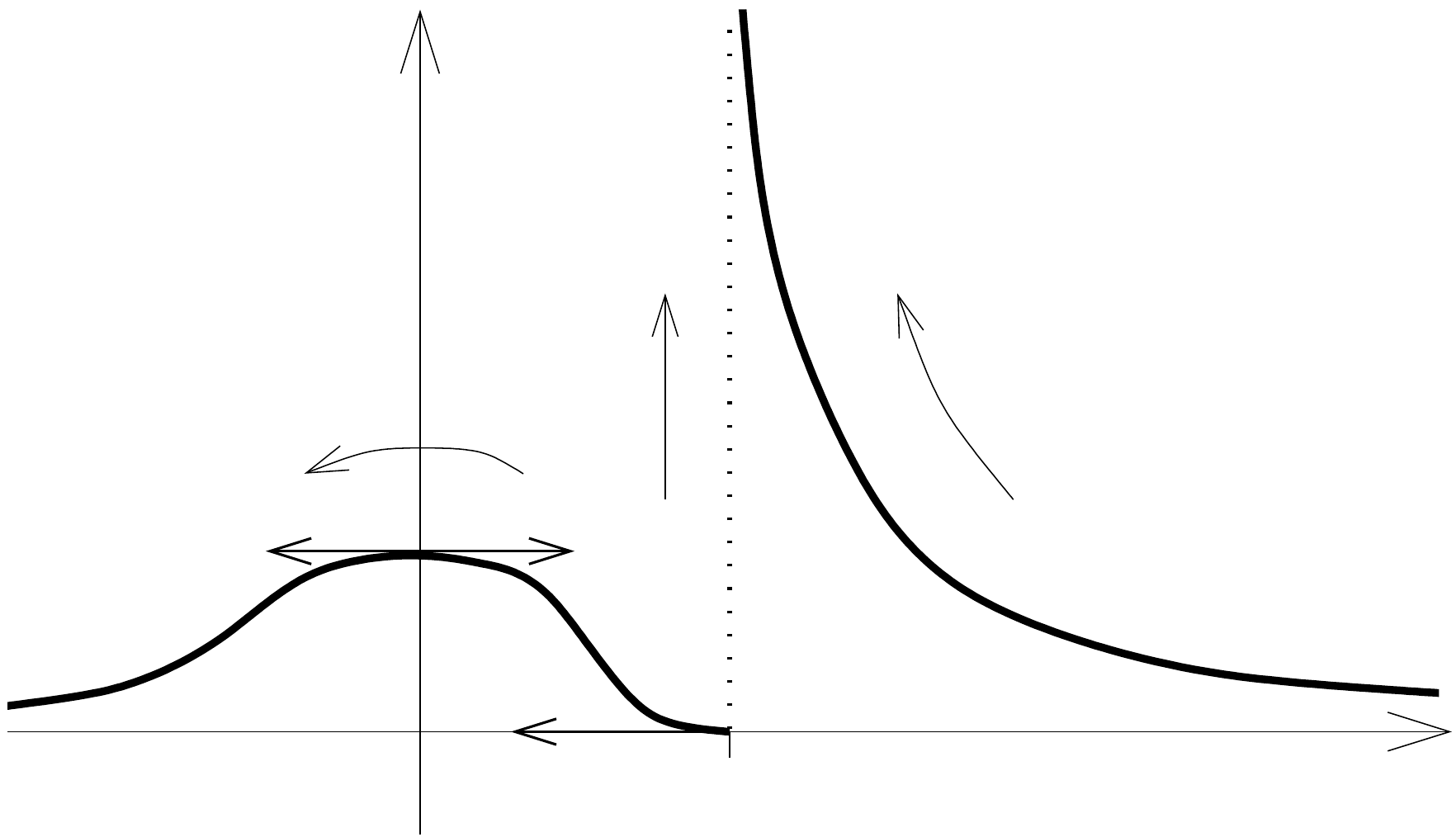}
\end{center}
\begin{picture}(0,0)
\put(146,27){$\tau$}\put(302,27){$\tau$}\put(458,27){$\tau$}
\put(35,106){$a$}\put(191,106){$a$}\put(347,106){$a$}
\put(3,76){$(i)$}\put(159,76){$(ii)$}\put(315,76){$(iii)$}
\put(221,27){$\tau_-$}\put(242,27){$\tau_+$}\put(389,26){$1$}
\end{picture}
\vspace{-.9cm}
\caption{\footnotesize \em Qualitative behaviours of the allowed branches of the scale factor $a$ as a function of time $\tau$: $(i)$~In the supercritical, $(ii)$ subcritical and $(iii)$ critical cases. The directions of the evolutions for increasing cosmic time $t$ are indicated for $c_\Phi>0$. Solid curves correspond to no-scale models with $\nF-\nB<0$. The dashed curve corresponds to no-scale models with $\nF-\nB>0$. Dotted lines correspond to super no-scale models, $\nF-\nB=0$.}
\label{fig_V}
\end{figure}

To be specific, integrating Eq.~(\ref{eq}), we find 
\be
\label{sol1}
a=a_0\,  {e^{-{1\over As }\arctan({\tau-1\over s  })}\over \P(\tau)^{1\over 2A}}\, 
\, , \qquad \where \qquad  s=\sqrt{1-\omega}\; \sqrt{\Big({c_\bot\over \gamma_cc_\Phi}\Big)^2-1} 
\ee
and $a_0$ is an integration constant, while combining this result with Eq.~(\ref{fri2}) yields 
\be
M^d= -{c_\Phi^2\over 2\alpha^2\omega\kappa^2}\, {a_0^{2A}\over (\nF-\nB)\, v_{d,n}}\,  {e^{-{2\over s}\arctan({\tau-1\over s})}\over a^{2(A+d-1)}}\, .
\label{m}
\ee
Fig.~\ref{fig_V}$(i)$ shows the qualitative shape of the scale factor as a function of $\tau$. The arrow shows the direction of the evolution for increasing cosmic time $t$, when $c_\Phi>0$. The opposite direction is realized by time-reversal, \ie by choosing  $c_\Phi<0$. We see that all solutions describe an initially growing universe that reaches a maximal  size before contracting. 

In the limits $\tau\to \epsilon \infty$, $\epsilon=\pm 1$, the expression $a(\tau)$ together with the definition of $\tau$ yield 
\be
a(t)\sim \left[{d(A+d-1)\over 2A}\, a_0^A \, e^{-{\epsilon \pi\over 2s}}\, \epsilon c_\Phi(t-t_\epsilon)\right]^{1\over A+d-1} ,
\ee
where $t_\epsilon$ is an integration constant. For $\epsilon c_\Phi>0$, this describes a Big Bang at $t\gtrsim t_\epsilon$, while $\epsilon c_\Phi<0$ corresponds to a Big Crunch at  $t\lesssim t_\epsilon$. Since $A>0$, we have in these regimes
\be
\label{assym}
H^2\sim\#\,  \dot\Phi^2\sim\#\,  \kappa^2\Vone\sim \#\, {a_0^{2A}c_\Phi^2\over a^{2(A+d-1)}}\gg {1\over 2}\, \dot\phi_\bot^2={c_\bot^2\over a^{2(d-1)}}\, ,
\ee
which shows that  the evolution of the universe at the Big Bang and Big Crunch is dominated by the no-scale modulus kinetic energy, partially compensated by the negative potential energy. As announced at the beginning of this section, the quantum effective potential plays also a fundamental role at the bounce, since 
\be
H=0 \quad \Longrightarrow\quad {1\over 2} \, (d-2)\, {\ddot a\over a}=-\K=\kappa^2\Vone<0\, .
\ee  

To study the domain of validity of perturbation theory during the cosmological evolution, it is enough to focus on the dilaton in the above $\tau\to \epsilon \infty$ limits. Eq.~(\ref{bot}) shows that asymptotically, $\phi_\bot$ converges to an integration constant, so that Eq.~(\ref{dil}) leads to
\be
\label{np}
e^{2d\alpha^2\phi}\sim \# \, \abs \tau\abs^{2\over \omega}\to +\infty\, .
\ee
Thus, the consistency of the 1-loop analysis is guaranteed late enough after the Big Bang and early enough before the Big Crunch. Moreover,  the scale factor is assumed to be large enough, for the kinetic energies in Eq.~(\ref{assym}) to be small compared to the string scale. This is required  not to have to take into account higher derivative terms in the effective action or, possibly, the dynamics of the whole string spectrum. For the above two reasons, the cosmological evolution can only be trusted far enough from its formal initial Big Bang ($t\gtrsim t_{\sign c_\Phi}$) and final Big Crunch ($t\lesssim t_{-\sign c_\Phi}$). 

To summarize, the supercritical case realizes a quantum universe whose existence is only allowed for a finite lapse of cosmic time (unless string theory resolves the Big Crunch and allows a never-ending evolution). Since the quantum corrections to the off-shell classical action allow new cosmological evolutions which describe the birth of a world sentenced to death, we may interpret this finite history as that of an ``unstable flat FLRW universe'' arising by quantum effects.  It is however not excluded that the expanding phase of the solution~(\ref{sol1}), (\ref{m}), (\ref{bot}) may be related in some way to some cosmological era of the real world.
 

\section{Subcritical case and quantum no-scale regimes}
\label{sous}

When the integration constants $c_\bot$  and $c_\Phi\neq 0$ satisfy the subcritical  condition
\be
 \left\abs{c_\bot\over c_\Phi}\right\abs<\gamma_c\, , 
\ee 
$\P(\tau)$ admits two distinct real roots,
\be
\tau_\pm=1\pm r\, , \qquad \where\qquad r=\sqrt{1-\omega}\; \sqrt{1-\Big({c_\bot\over \gamma_cc_\Phi}\Big)^2}\, .
\ee
Important remarks follow from Friedmann  equation~(\ref{fri2}). First, the bosonic or fermionic nature of the massless spectrum determines the allowed ranges of variation of $\tau$, 
\begin{align}
\nF-\nB <0\qquad &\Longrightarrow \qquad \tau<\tau_- \quad \mbox{or} \quad \tau>\tau_+\nonumber \\
\nF-\nB =0\qquad &\Longrightarrow \qquad \tau\equiv \tau_- \quad \mbox{or}\quad \tau\equiv \tau_+\nonumber\\
\nF-\nB >0\qquad &\Longrightarrow \qquad \tau_-<\tau<\tau_+\, .\label{inter}
\end{align}
Second, taking the classical limit $\kappa^2\to 0$ is allowed, and yields evolutions $\tau(t)\equiv \tau_-$ or $\tau(t)\equiv \tau_+$. Therefore, the classical trajectories are  identical to those obtained for quantum super no-scale models, \ie when $\nF-\nB=0$. In the following, we start by describing the cosmological solutions in the super no-scale case, and then show that the quantum evolutions for generic no-scale models ($\nF-\nB\neq 0$) admit quantum no-scale regimes, \ie behave  the same way. 


\subsection{Case \bm  $\nF-\nB=0$}

When $\tau\equiv \tau_\pm$,\footnote{One may think that the space of solutions in the super no-scale case is divided in two parts, corresponding to either $\tau\equiv \tau_+$ or  $\tau\equiv \tau_-$. This is however not true. Including the critical case of Sect.\ref{cri2}, all the evolutions are actually of the form $\tau\equiv \tau_i$, where $1-\sqrt{1-\omega}\le \tau_i\le 1+\sqrt{1-\omega}$.} Eqs~(\ref{eq}) and~(\ref{fri2}) being trivial, we use the definition of $\tau$ given in Eq.~(\ref{tau}) to derive the scale factor as a function of cosmic time $t$, 
\be
a= \left[{d(d-1)\over 2A} \, (1\pm r) \, c_\Phi(t-t_\pm)\right]^{1\over d-1}, 
\ee
where $t_\pm$ is an  integration constant. For $c_\Phi>0$, this describes a never-ending era $t>t_\pm$ of expansion, initiated by a Big Bang occurring at $t=t_\pm$. Of course, the solution obtained by time-reversal satisfies $c_\Phi<0$ and describes an era $t<t_\pm$ of  contraction that ends at the Big Crunch occurring at $t_\pm$. Integrating the no-scale modulus equation in~(\ref{cc}), we find 
\be
M^d={e^{d\alpha\Phi_{\pm}}\over a^{2(d-1)+K_\pm}}\, \Ms^d\, , 
\ee
where $\Phi_\pm$ is an  integration constant and 
\be
\label{k}
K_\pm=\pm {2Ar\over 1\pm r}\, .
\ee
In total, when the 1-loop effective potential vanishes (up to exponentially suppressed terms), the cosmological evolution is driven by the kinetic energies of the free scalar fields, 
\be
H^2\propto \dot\phi_\bot^2  \propto  \dot\Phi^2\propto {c_\Phi^2\over a^{2(d-1)}}\, .
\ee 

The dilaton evolution is found using Eq. (\ref{dil}),
\be
e^{2d\alpha^2\phi}= {e^{d\alpha \Phi_\pm}\, e^{ {d\over n}\sqrt{d-2+n}\, \phi_{\bot \scriptscriptstyle \pm}}\over a^{P_\pm}}\, ,
\ee
where $\phi_{\bot \scriptscriptstyle \pm}$ is an integration constant and 
\be
\label{p+-}
P_\pm={K_\pm\over \omega r}\Big(r\pm \big(1-\omega-{\omega\over n}\sqrt{2(d-2+n)}\; {c_\bot\over c_\Phi}\big)\Big)\,.
\ee
Unless $P_\pm$ vanishes, in which case the dilaton is constant, the string coupling $\gs=e^{\phi}$ varies monotonically between perturbative and non-perturbative regimes. For instance, the solution $\tau\equiv \tau_+$ is perturbative in the large scale factor limit,  $c_\Phi (t-t_+)\to +\infty$, when $P_+>0$. In order to translate this condition into a range for $c_\bot/ c_\Phi$, we introduce 
\be
\gamma_\pm = {(1-\omega)\sqrt{2(d-2)\over n}\pm \alpha\sqrt{2(1-\omega)(d-2)}\over 2\alpha(\omega {d-2\over n}+1)}\, , 
\ee
which satisfy $0<\pm\, \gamma_\pm <\gamma_c$, and find $P_+>0$ if and only if 
\be
-\gamma_c< {c_\bot\over c_\Phi}<\gamma_+ \mbox{ \, when \,  }n<{d^2(d-2)\over 4(d-1)}\, , \qquad 
-\gamma_c< {c_\bot\over c_\Phi}<\gamma_c\mbox{ \, when \,  } n>{d^2(d-2)\over 4(d-1)}\, .
\label{1}
\ee
In a similar way, the solution $\tau\equiv \tau_-$ is perturbative in the small scale factor limit, $c_\Phi (t-t_-)\to 0_+$, when $P_-<0$ \ie
\be
\gamma_-< {c_\bot\over c_\Phi}<\gamma_c \mbox{ \, when \,  }n<{d^2(d-2)\over 4(d-1)}\, , \qquad 
\gamma_-< {c_\bot\over c_\Phi}<\gamma_+\mbox{ \, when \,  } n>{d^2(d-2)\over 4(d-1)}\, .
\label{2}
\ee
As already mentioned in the supercritical case, beside the conditions for the $\gs$-expansion to be valid, the above solutions suppose the scale factor to be large enough, for the higher derivative terms ($\alpha'$-corrections) to be small. Because of this constraint, the Big Bang ($t\gtrsim t_\pm$) and Big Crunch ($t\lesssim t_\pm$) behaviours are only formal. 

\subsection{Case \bm $\nF-\nB \neq 0$}

\noindent Let us turn to the analysis of a generic no-scale model, thus characterized by $\nF-\nB\neq 0$. In this case, $\tau$ can actually be treated as a time variable and, integrating Eq.~(\ref{eq}), we find
\be
\label{sol}
a={a_0\over \abs \tau-\tau_-\abs^{1\over K_-}\, \abs\tau-\tau_+\abs^{1\over K_+}}\, , 
\ee
where $a_0>0$ is an integration constant. Using this result with  Friedmann equation~(\ref{fri2}) yields $M^d$, which can be written in two different suggestive ways,  
\be
\label{fri3}
M^d={M^d_+\over a^{2(d-1)+K_+}} \left\abs{\tau-\tau_-\over \tau_+-\tau_-}\right\abs^{2\over \tau_+}={M^d_-\over a^{2(d-1)+K_-}} \left\abs{\tau_+-\tau\over \tau_+-\tau_-}\right\abs^{2\over \tau_-} , 
\ee
where we have defined 
\be
\label{M+-}
M_\pm^d=  {c_\Phi^2\over 2\alpha^2\omega\kappa^2}\, {a_0^{K_\pm}\over \abs \nF-\nB\abs \,  v_{d,n}}\, \abs\tau_+-\tau_-\abs^{2\over \tau_\pm}\, .
\ee
Fig.~\ref{fig_V}$(ii)$ shows schematically the scale factor $a$ as a function of $\tau$. As expected from Eq.~(\ref{inter}), two branches (in solid lines) exist when $\nF-\nB<0$. Thus, depending on the choice of initial condition $\tau_i\equiv \tau(t_i)$, the scale factor evolves along one or the other. When $\nF-\nB>0$, a single branch (in dashed line) can be followed by the universe. For completeness, the constant $\tau\equiv \tau_\pm$ trajectories found in the previous subsection for the super no-scale case $\nF-\nB=0$ are also displayed (in dotted lines). The arrows indicate the directions of the evolutions for increasing cosmic time $t$, when $c_\Phi>0$. The opposite directions are realized by time-reversal, with $c_\Phi<0$.  We see that all branches start and/or end with a vanishing scale factor, when $\tau\to \pm \infty$ or $\tau\to \tau_-$.\footnote{The trajectories allowing $\tau$ to approach $\tau_-$ are shown in Fig.~\ref{fig_V}$(ii)$ in the case $da/d\tau\to \pm \infty$, when $\tau\to \tau_-$. This occurs when $\abs c_\bot/c_\Phi\abs<\gamma_M$, where $\gamma_M=[ ({1\over\omega}(1-{1\over (1+2A)^2})-1) {1\over 2\alpha^2}]^{1\over 2}$. On the contrary, when $\gamma_M<\abs c_\bot/c_\Phi\abs<\gamma_c$, one has $da/d\tau=0$ at $\tau=\tau_-$. Finally, $\abs c_\bot/c_\Phi\abs = \gamma_M$ implies $\abs da/d\tau\abs$ to be finite and non-vanishing  at $\tau=\tau_-$. However, these different behaviours do not play any important role in the sequel.} In all cases, whether $da/d\tau$ vanishes, is infinite or is finite when $a(\tau)\to 0$, we will see that $da/dt$  diverges at a finite cosmic time, thus describing a formal Big Bang or Big Crunch. 

Note that when $\nF-\nB\neq 0$, all branches allow $\tau$ to approach $\tau_+$ and/or $\tau_-$. When this is the case, the behaviour $\tau(t)\to \tau_\pm$ yields, using the definition of $\tau$ given in Eq.~(\ref{tau}),
\be
\label{t}
a\sim \left[{d(d-1)\over 2A} \, (1\pm r) \, c_\Phi(t-t_\pm)\right]^{1\over d-1}, \qquad 
M^d\sim{M^d_\pm\over a^{2(d-1)+K_\pm}}\, ,
\ee
for some integration constant $t_\pm$.  
This shows that the cosmological evolution of the universe as well as that of the scalars $\Phi$ and $\phi_\bot$ approach those found in the super no-scale case $\nF-\nB=0$, \ie for vanishing 1-loop effective potential (up to exponentially suppressed terms). For this reason, {\em we  define the limits $\tau\to \tau_\pm$ of the generic no-scale models as ``quantum no-scale regimes''.} These are characterized by phases of the universe dominated by the scalar kinetic energies,
\be
H^2\sim \# \, \dot \phi_\bot^2\sim \#\,  \dot\Phi^2\sim \#\, {c_\Phi^2\over a^{2(d-1)}}\gg \kappa^2\abs \Vone\abs \sim \#\, {a_0^{K_\pm}c_\Phi^2\over a^{2(d-1)+K_\pm}}\, .
\ee
In fact, when $\tau\to \tau_+$,  the divergence of  the scale factor, $a(\tau)\to+\infty$, and the fact that $K_+>0$ imply that the quantum potential is effectively dominated. Moreover, Eq.~(\ref{t})   shows that this regime lasts for an indefinitely long cosmic time, $c_\Phi t\to +\infty$. In a similar way, when  $\tau\to \tau_-$, since $a(\tau)\to 0$ and $K_-<0$, the effective potential is again dominated, and this process is realized when cosmic time approaches $t_-$, $\mbox{$c_\Phi (t-t_-)\to 0_+$}$. To summarize, when $\tau\to \tau_\pm$, assuming a perturbative regime, the quantum cosmological evolution of the no-scale model is attracted to that of a classical background ($\kappa^2=0$), where the no-scale structure is exact. In particular, the temporal evolution of the no-scale modulus $\Phi$ approaches that of a free field, so that  the no-scale structure tends to withstand perturbative corrections. Note however that the cosmological solutions found for no-scale models satisfying $\nF-\nB<0$ can also admit other regimes. As in the supercritical case, the latter correspond to limits $\tau\to \epsilon \infty$, $\epsilon=\pm 1$, describing a formal Big Bang or Big Crunch,
\be
a(t)\sim \left[{d(A+d-1)\over 2A}\, a_0^A\, \epsilon c_\Phi(t-t_\epsilon)\right]^{1\over A+d-1} , 
\label{de}
\ee
where $t_\epsilon$ is an integration constant. In these circonstances, the universe is dominated by the kinetic energy and negative potential of the no-scale modulus, as summarized in Eq.~(\ref{assym}). 

To determine the domain of validity of perturbation theory, we integrate Eq.~(\ref{bot}), which introduces an arbitrary  constant mode $\phi_{\bot 0}$, and use Eq. (\ref{dil}) to derive the time dependance of the dilaton,
\be
e^{2d\alpha^2\phi}= {c_\Phi^2\over 2\alpha^2\omega\kappa^2\Ms^d}\, {1\over \abs \nF-\nB\abs \, v_{d,n}}\, {e^{ {d\over n}\sqrt{d-2+n}\, \phi_{\bot 0}}\over a_0^{2(d-1)}}\; \abs \tau-\tau_+\abs^{L_+}\;  \abs \tau-\tau_-\abs^{L_-} \, , \;\;\,\quad L_\pm={P_\pm\over K_\pm}\, , 
\ee
where $P_\pm$ is defined in Eq.~(\ref{p+-}). Therefore, the conditions $L_\pm>0$ for perturbative consistency of the attractions to the quantum no-scale regimes $\tau\to \tau_\pm$ are those found in the super no-scale case: For $\tau\to \tau_+$, $c_\bot/c_\Phi$ must satisfy Eq.~(\ref{1}), while for $\tau\to \tau_-$,  $c_\bot/c_\Phi$ must respect Eq.~(\ref{2}). In particular,  when $\nF-\nB>0$, the cosmological evolution between $\tau_-$ and $\tau_+$ is all the way perturbative if $\gamma_-<c_\bot/c_\Phi<\gamma_+$. In this case, the quantum potential is negligible, $\Vone\ll \K$, throughout the evolution, except in the vicinity of $\tau= 1$, where it induces the transition from one no-scale regime to the other. On the contrary, the regimes $\tau\to \epsilon\infty$, which can be reached when $\nF-\nB<0$, can be trusted up to the times the evolutions become non-perturbative, as follows from Eq.~(\ref{np}), which is valid for arbitrary $c_\bot/c_\phi\in \R$.  

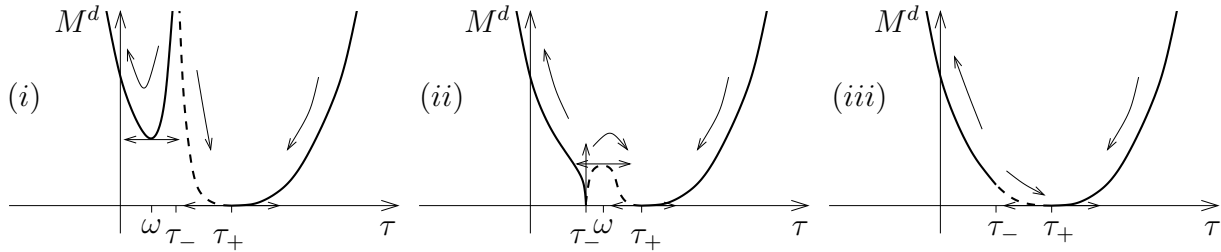
\begin{figure}[h!]
\vspace{.6cm}
\begin{center}
\includegraphics[height=3.0cm]{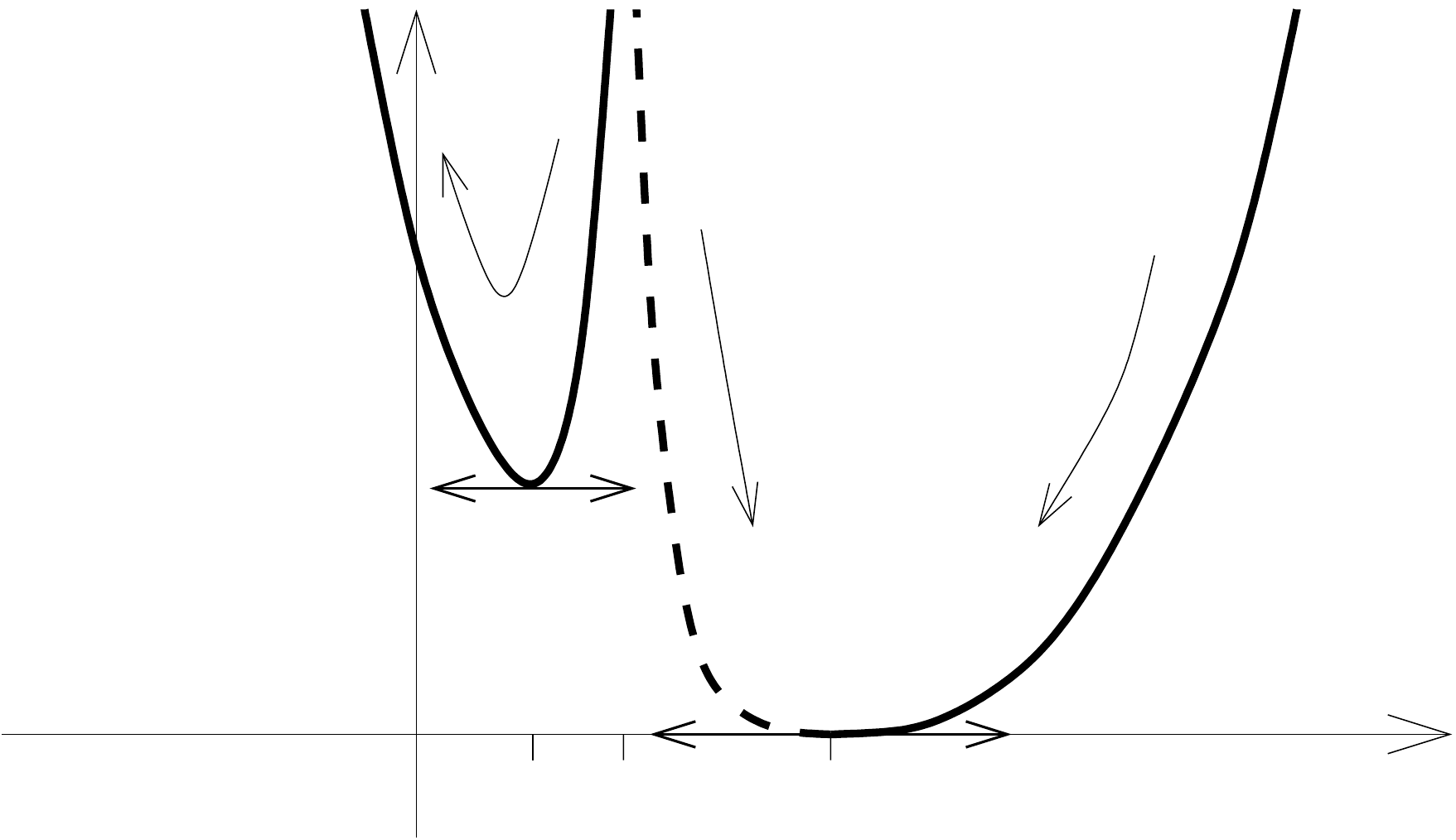}\;
\includegraphics[height=3.0cm]{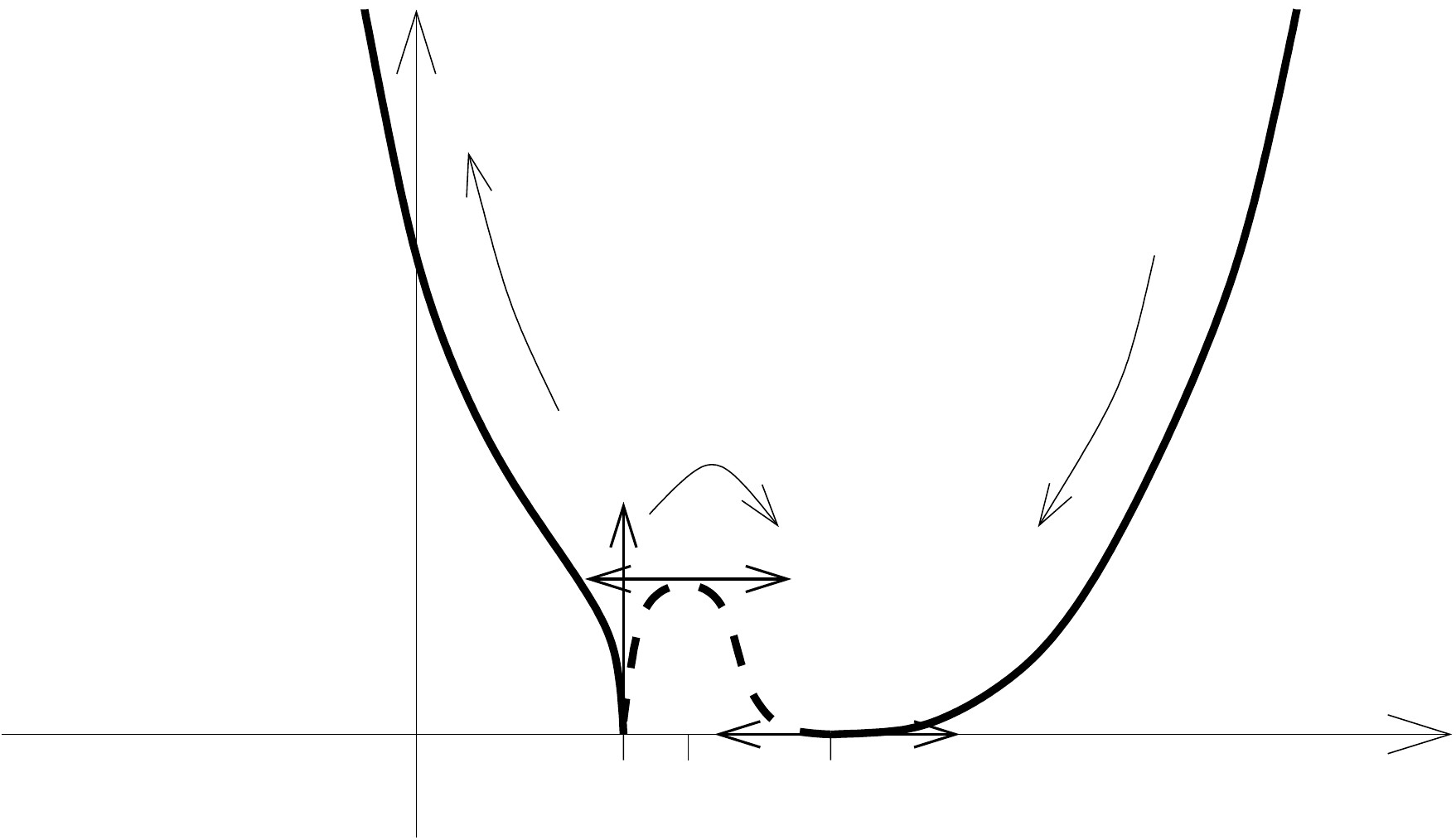}\;
\includegraphics[height=3.0cm]{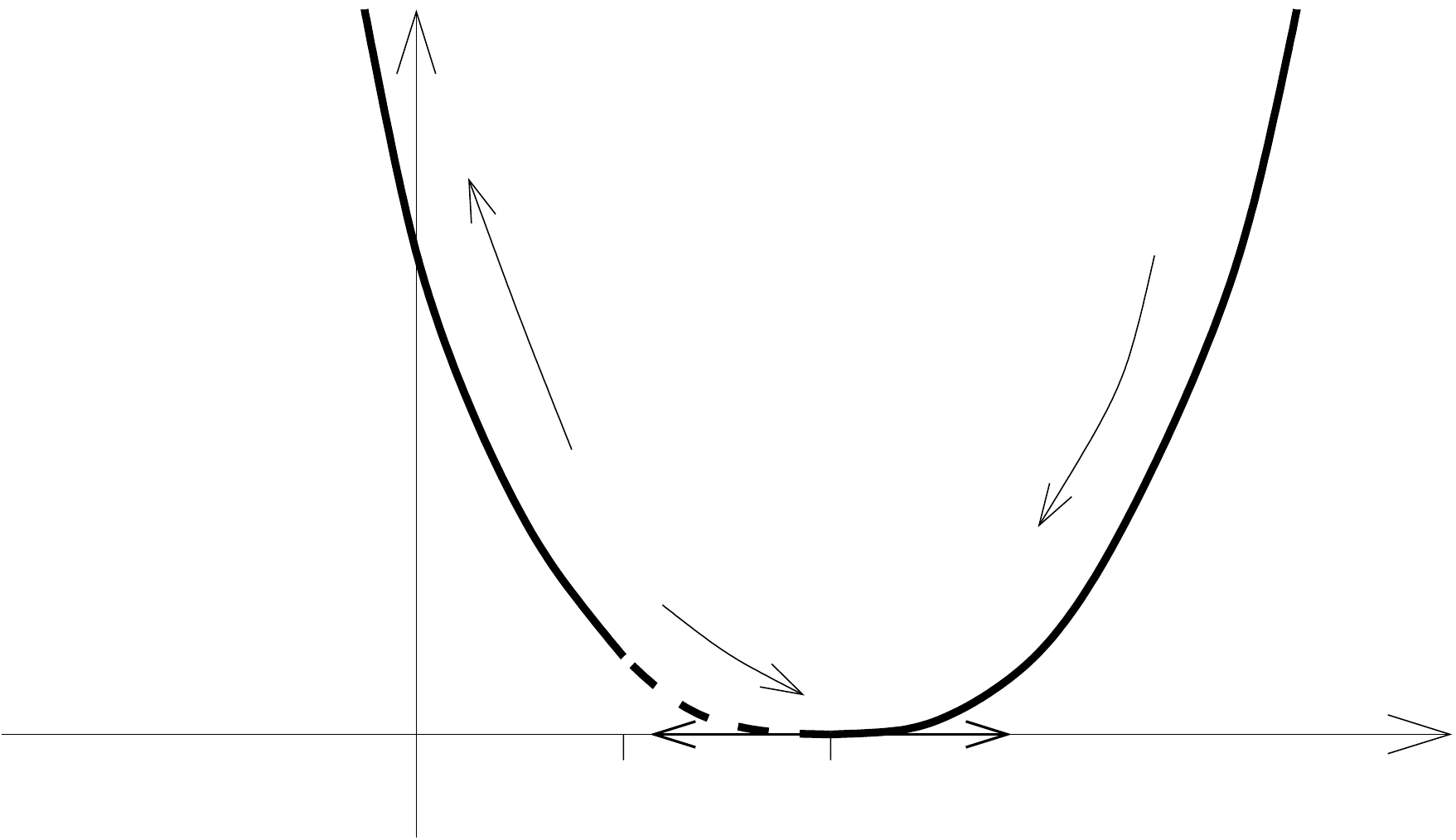}
\end{center}
\begin{picture}(0,0)
\put(146,27){$\tau$}\put(302,27){$\tau$}\put(458,27){$\tau$}
\put(23,104){$M^d$}\put(178,104){$M^d$}\put(333,104){$M^d$}
\put(5,76){$(i)$}\put(161,76){$(ii)$}\put(317,76){$(iii)$}
\put(56,28){$\omega$}\put(65,24){$\tau_-$}\put(83,24){$\tau_+$}\put(218,24){$\tau_-$}\put(227,28){$\omega$}\put(241,24){$\tau_+$}\put(376,27){$\tau_-$}\put(397,27){$\tau_+$}
\end{picture}
\vspace{-.9cm}
\caption{\footnotesize \em Qualitative behaviours of the allowed branches of $M^d$ as a function of $\tau$, in the subcritical case: $(i)$ When $\sqrt{\omega}\gamma_c<\abs c_\bot/c_\Phi\abs<\gamma_c$, $(ii)$ when $\abs c_\bot/c_\Phi\abs<\sqrt{\omega}\gamma_c$, $(iii)$ or when $\abs c_\bot/c_\Phi\abs=\sqrt{\omega}\gamma_c$. The directions of the evolutions for increasing cosmic time $t$ are indicated for $c_\Phi>0$. Solid curves refer to no-scale models with $\nF-\nB<0$, while the dashed ones refer to no-scales models with $\nF-\nB>0$.}
\label{fig_Md}
\end{figure}

For $\nF-\nB\neq 0$, the subcritical case can also give rise to non-trivial dynamics of the no-scale modulus, which can be summarized as follows, for instance in the case $c_\Phi>0$: 

$(i)$ When $\sqrt{\omega} \gamma_c<\abs c_\bot/c_\Phi\abs < \gamma_c$,  even if the effective potential is dominated in the no-scale regime $\tau\to\tau_-$, $M^d$ turns out  to diverge, as can be seen in Fig.~\ref{fig_Md}$(i)$, which shows the three branches $M^d$ can follow as a function of $\tau$. The directions of the evolutions for increasing cosmic time $t$ are again indicated for $c_\Phi>0$. Along the trajectories satisfying $\tau>\tau_+$, the universe expands and is attracted to the no-scale regime $\tau\to \tau_+$, while $M(t)$ decreases. Thus, even if this is counterintuitive, the supersymmetry breaking scale forever climbs its negative effective potential $(\nF-\nB)v_{d,n}M^d$ \cite{DS}. This fact contradicts the expectation that $M$ should increase  and yield a large, negative potential energy.
On the contrary, the situation is more natural in the other branches. For the solutions satisfying $\tau<\tau_-$, if $M(t)$ also starts climbing its negative potential, it is afterwards attracted back to large values, with the turning point sitting at $\tau=\omega$. Finally, along the branch $\tau_-<\tau<\tau_+$, the universe expands and is attracted to the no-scale regime $\tau\to \tau_+$, while $M(t)$ drops along its positive potential $(\nF-\nB)v_{d,n}M^d$. 

$(ii)$ When $\abs c_\bot/c_\Phi\abs < \sqrt{\omega} \gamma_c$,  as shown in Fig.~\ref{fig_Md}$(ii)$, $M^d$ vanishes when $\tau\to \tau_-$. Along the branch $\tau>\tau_+$, $M(t)$ climbs as before its negative potential \cite{DS}, while for $\tau<\tau_-$, it drops. The branch $\tau_-<\tau<\tau_+$ is the most interesting one:  While the scale factor increases, $M(t)$ first climbs the positive potential $(\nF-\nB)v_{d,n}M^d$, and then descends. At the turning point located at $\tau=\omega$, we have $\dot \Phi=0$ and $\Vone>0$, which is enough to show that for small enough  $\abs c_\bot\abs$,  the scale factor accelerates for a lapse of cosmic time \cite{solV, DS, accel}. However, the resulting $e$-fold number being of order 1, this acceleration of the universe does not yield any efficient inflationary effect\footnote{We warmly thank Lucien Heurtier to have analyzed the magnitude of the $e$-fold number.}.

$(iii)$ Finally, Fig.~\ref{fig_Md}$(iii)$ plots $M^d(\tau)$ in the limit case $\abs c_\bot/c_\Phi\abs=\sqrt{\omega}\gamma_c$. 

To conclude on the subcritical case, let us stress again that the dynamics arising from initial conditions such that
\begin{align}
c_\Phi>0 \;\;\when \;\;\nF-\nB\ge 0\, , \qquad  \mbox{or}\qquad c_\Phi>0 \;\;\and \;\; \tau_i>\tau_+ \;\;\when \;\;\nF-\nB<0\, , 
\end{align}
and satisfying Eq.~(\ref{1}) to insure validity of perturbation theory, forces the universe to enter the no-scale regime $\tau(t)\to \tau_+$.\footnote{At 1-loop, the super no-scale models, $\nF-\nB=0$, also admit the expanding solution $c_\phi>0$, $\tau(t)\equiv \tau_-$, which is perturbative if $\abs c_\bot/c_\Phi\abs <\gamma_c$ satisfies the condition complementary to that given in Eq.~(\ref{2}). However, we expect this statement to be invalidated when higher order corrections in $\gs$ are taken into account and the effective potential is no more vanishing (up to exponentially suppressed terms).}    After the Big Bang, the solutions describe flat, ever-expanding FLRW universes that are not destabilized by quantum effects. Instead of generating a cosmological constant, quantum corrections induce a time-dependent effective potential proportional to $M^d$ and asymptotically negligible, from a cosmological point of view. Note however that $M$ plays a fundamental role in the effective $d$-dimensional renormalizable theory in rigid Minkowski space, which is found by keeping only the relevant operators present in supergravity. In fact, $M$ determines the order of magnitude of all soft breaking terms in the resulting MSSM-like theory of particles \cite{Barbieri:1982eh}. 


\section{Critical case}
\label{cri}

In the critical case, which corresponds to 
\be
\left\abs{c_\bot\over c_\Phi}\right\abs=\gamma_c\, , 
\ee 
the polynomial $\P(\tau)$ has a double root $\tau_+=\tau_-=1$, and the no-scale model must satisfy $\nF-\nB\leq 0$, as follows from Eq.~(\ref{fri2}). In the sequel, we show that  the qualitative  behaviour of the associated cosmological evolutions are similar to those found in the subcritical case. 


\subsection{Case \bm  $\nF-\nB=0$}
\label{cri2}

The  solutions in the super no-scale case are actually those found in the subcritical case for $c_\bot/c_\Phi=\eta\gamma_c$, $\eta=\pm 1$, \ie for $r=0$: 
\be
a= \left[{d(d-1)\over 2A} \, c_\Phi(t-t_0)\right]^{1\over d-1} , \quad M^d={e^{d\alpha\Phi_0}\over a^{2(d-1)}}\, \Ms^d\, ,\quad  e^{2d\alpha^2\phi}= {e^{d\alpha \Phi_0}\, e^{ {d\over n}\sqrt{d-2+n}\, \phi_{\bot 0}}\over a^{P_0}}\, ,
\ee
where $t_0$, $\Phi_0$, $\phi_{\bot0}$ are integration constants and 
\be
\label{p0}
P_0=2(d-1)\!\left(1-\eta \sqrt{{d-2\over n}\, {\omega\over 1-\omega}}\right). 
\ee
The above evolutions are  perturbative in the large scale factor regime, $c_\Phi (t-t_0)\to +\infty$, when $P_0>0$. This is the case for $c_\bot/c_\Phi = -\gamma_c$, as well as for $c_\bot/c_\Phi=\gamma_c$ if $n>{d^2(d-2)\over 4(d-1)}$. 


\subsection{Case \bm  $\nF-\nB<0$}

For the no-scale models with negative 1-loop effective potentials, Eqs~(\ref{eq}) and~(\ref{fri2}) yield  
\be
a=a_0\, {e^{1\over A(\tau-1)}\over \abs \tau-1\abs^{1\over A}}\, , \qquad M^d= {c_\Phi^2\over 2\alpha^2\omega\kappa^2}\, {-1\over ( \nF-\nB) \,  v_{d,n}}\, {(\tau-1)^2\over a^{2(d-1)}}\, ,
\ee
where $a_0>0$ is an integration constant. 
Fig.~\ref{fig_V}$(iii)$ shows in solid lines the two branches $\tau>1$ and $\tau<1$ the scale factor $a(\tau)$ can follow, while the dotted line $\tau\equiv 1$ corresponds to the critical super no-scale case. The trajectories admit one of the two limits $\tau\to 1_+$ or $\tau\to 1_-$, which lead in terms of cosmic time to
\be
\label{t1}
a\sim \left[{d(d-1)\over 2A} \, c_\Phi(t-t_\pm)\right]^{1\over d-1},
\ee
where $t_\pm$ is an integration constant.
The behaviour $\tau\to 1_+$ describes an expanding or contracting flat FLRW solution, $c_\Phi (t-t_+)\to +\infty$, while $\tau\to 1_-$ corresponds to a Big Bang or Big Crunch, $c_\Phi(t-t_-)\to 0_+$.
Note that even if $K_\pm$ in Eq.~(\ref{k}) vanishes for $r=0$, the effective potential is still dominated by the moduli kinetic energies, 
\be
H^2\sim \# \, \dot \phi_\bot^2\sim \#\,  \dot\Phi^2\sim \#\, {c_\Phi^2\over a^{2(d-1)}}\gg \kappa^2\abs \Vone\abs \sim \#\, c_\Phi^2{(\tau-1)^2\over a^{2(d-1)}}\, ,
\ee
which proves that the limits $\tau\to 1_\pm$ describe quantum no-scale regimes. 
On the contrary, the limits $\tau\to \epsilon\infty$, $\epsilon=\pm 1$, yield Big Bang/Big Crunch behaviours, as shown in Eqs~(\ref{de}) and~(\ref{assym}). 

Finally, for $c_\bot/c_\Phi=\eta\gamma_c$, $\eta=\pm 1$,  the dilaton trajectory is given by
\be
e^{2d\alpha^2\phi}= {c_\Phi^2\over 2\alpha^2\omega\kappa^2\Ms^d}\, {-1\over ( \nF-\nB) \, v_{d,n}}\, {e^{ {d\over n}\sqrt{d-2+n}\, \phi_{\bot 0}}\over a_0^{2(d-1)}}\; (\tau-1)^{2\over \omega}\, e^{-{P_0\over A(\tau-1)}}\, ,
\ee
where $P_0$ is defined in Eq.~(\ref{p0}) and $\phi_{\bot 0}$ is the arbitrary  constant mode of $\phi_\bot$.   
Thus,\footnote{These remarks can be recovered from Eqs~(\ref{1}) and~(\ref{2}) by taking $c_\bot/c_\Phi\to \eta \gamma_c$} the no-scale regime $\tau\to 1_+$  is perturbative for $c_\bot/c_\Phi=-\gamma_c$, as well as for $c_\bot/c_\Phi=\gamma_c$ if $n>{d^2(d-2)\over 4(d-1)}$. On the contrary, the regime  $\tau\to 1_-$ is perturbative only for $c_\bot/c_\Phi=\gamma_c$, if $n<{d^2(d-2)\over 4(d-1)}$. In the limits $\tau\to \epsilon\infty$, $\epsilon=\pm 1$, the models are non-perturbative (see Eq. (\ref{np})). 


\section{\bm Case $c_\Phi=0$}
\label{part}

What remains to be presented is the cosmological evolution for $c_\Phi=0$. The supersymmetry breaking scale can be found by integrating the no-scale modulus equation in~(\ref{cc}), which gives 
\be
M^d={e^{d\alpha\Phi_0}\over a^{2(A+d-1)}}\, \Ms^d\, , 
\ee
where $\Phi_0$ is a constant. This result can be used to write Friedmann equation~(\ref{fri}) as
\be
{A\over 2}\, (d-2)\, H^2=-{c_\bot^2\over a^{2(d-1)}}-(\nF-\nB)\, v_{d,n}\, {e^{d\alpha\Phi_0}\over a^{2(A+d-1)}}\, \kappa^2\Ms^d\, ,
\ee
which for $c_\bot\neq 0$ requires the no-scale model to satisfy $\nF-\nB<0$. Note that this is not surprising since this case is somehow ``infinitely supercritical''. As a result, no-scale regimes are not expected to exist. To be specific, when $c_\bot\neq 0$,  the above differential equation can be used to determine  the cosmic time as a function of the scale factor, 
\begin{align}
t&=t_*-\epsilon\,  {a_{\rm max}^{A+d-1}\over c_v}\int^1_{a/a_{\rm max}}{x^{A+d-2}\, dx\over\sqrt{1-x^{2A}}}\, ,  \qquad \where \qquad \epsilon=\pm 1\, ,  \nonumber \\
 a_{\rm max}^{2A}&={\abs \nF-\nB\abs\, v_{d,n}\, e^{d\alpha\Phi_0} \over c_\bot^2}\, \kappa^2\Ms^d \,,  \qquad c_v^2={2\over d-2}\, {\abs \nF-\nB\abs\, v_{d,n}\,e^{d\alpha\Phi_0}\over A}\, \kappa^2\Ms^d\, ,\esp 
\end{align}
and $t_*$ is an integration constant. Defining
\be
t_\epsilon=t_*-\epsilon t_0\, , \qquad t_0={a_{\rm max}^{A+d-1}\over c_v}\int^1_0{x^{A+d-2}\, dx\over\sqrt{1-x^{2A}}} \, , 
\ee
the time variable $t$ varies from $t_+$ to $t_-$. At $t=t_+$, a Big Bang initiates an era of expansion that stops when the scale factor reaches its maximum $a_{\rm max}$ at $t=t_*$. Then, the universe contracts until a Big Crunch occurs at $t=t_-$. Close to the initial and final times $t_\pm$, the scale factor behaves as 
\be
a(t)\sim \big(c_v\, (A+d-1)\, \epsilon (t- t_\epsilon)\big)^{1\over A+d-1} , 
\label{ss}
\ee
which leads to scalings similar to those given in Eq. (\ref{assym}), namely
\be
\label{assym2}
H^2\sim\#\,  \dot\Phi^2\sim\#\,  \kappa^2\Vone\sim \#\, {c_v^2\over a^{2(A+d-1)}}\gg {1\over 2}\, \dot\phi_\bot^2={c_\bot^2\over a^{2(d-1)}}\, .
\ee
However, the scalar $\phi_\bot$ converges in this case to a constant, so that  Eq.~(\ref{dil}) yields
\be
e^{2d\alpha^2\phi}\sim {\# \over c_v^2(t- t_\epsilon)^2}\to +\infty\, .
\label{p}
\ee
As a result of Eqs~(\ref{assym2}) and~(\ref{p}), the cosmological solution we have found can only be trusted far enough from the formal Big Bang and Big Crunch, not to have to take into account $\alpha'$- and $\gs$-corrections. 

Finally, when $c_\bot=0$ and $\nF-\nB<0$, the maximum scale factor $a_{\rm max}$ is formally infinite, so that no turning point exists anymore.  In fact, relations~(\ref{ss}) and~(\ref{p}) become equalities: The evolution for $\epsilon =+1$ describes a never-ending era $t>t_+$ of expansion, while the trajectory for $\epsilon =-1$ describes an era $t<t_-$ of  contraction. For super no-scale models, \ie when $\nF-\nB=0$, the case $c_\bot=0$ yields the trivial solution where all fields $a$, $\Phi$, $\phi_\bot$ are static. 


\section{Summary and conclusion}
\label{cl}

\begin{figure}[h!]
\vspace{.6cm}
\begin{center}
\includegraphics[height=6.0cm]{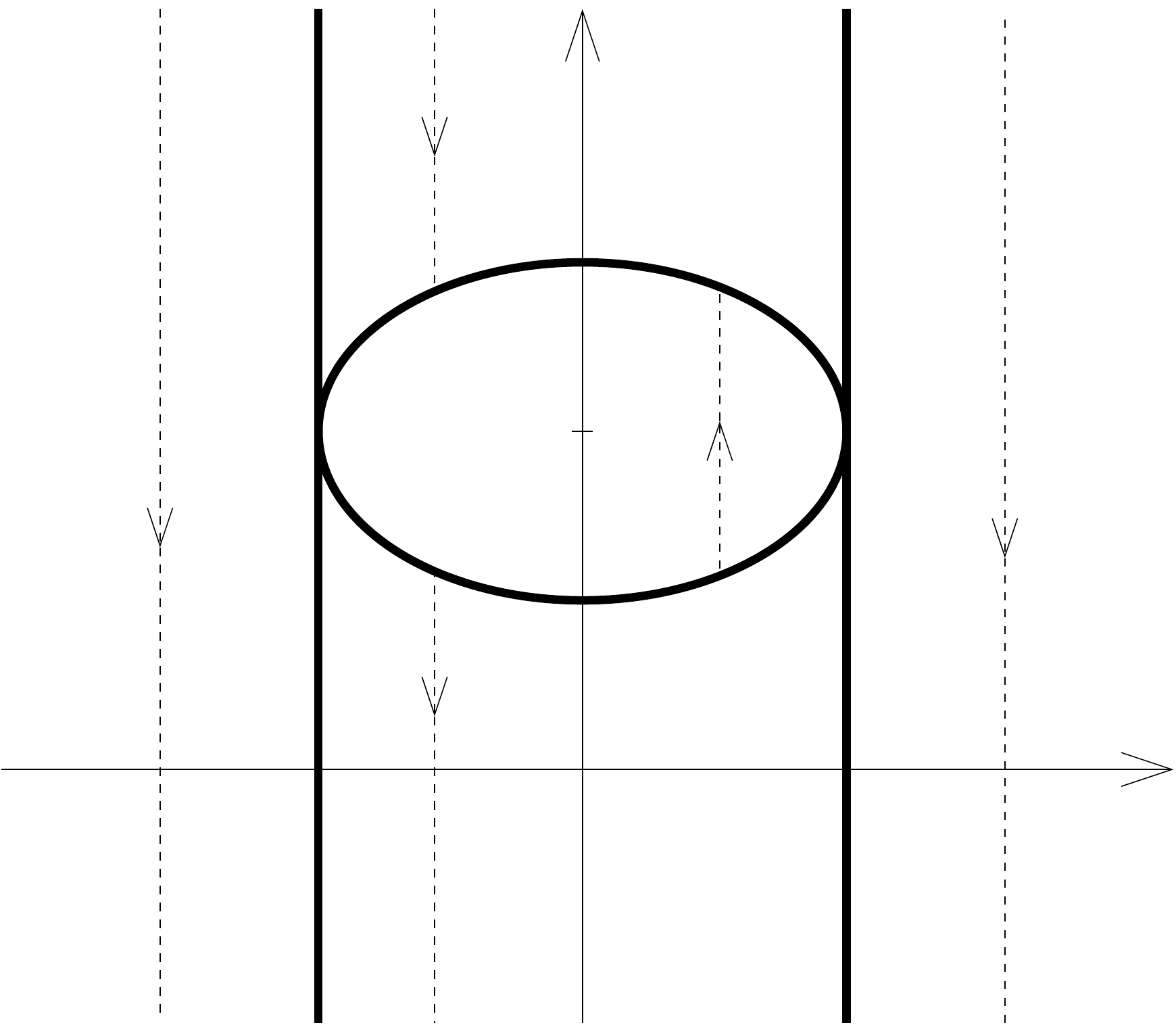}
\end{center}
\begin{picture}(0,0)
\put(319,50){$\displaystyle {c_\bot\over c_\Phi}$}
\put(219,191){$\tau_i$}
\put(235,157.5){$\scriptscriptstyle 1+\sqrt{1-\omega}$}
\put(237,123){$\textstyle  1$}
\put(235,92){$\scriptscriptstyle  1-\sqrt{1-\omega}$}
\put(280,60){$\textstyle  \gamma_c$}
\put(168.5,60){$\textstyle  -\gamma_c$}
\put(311,123){(I)}
\put(140,123){(I$'$)}
\put(240,173){(II)}
\put(205,123){(III)}
\put(240,43){(IV)}
\end{picture}
\vspace{-.9cm}
\caption{\footnotesize \em Partition of the $\R^2$-plane $(c_\bot/c_\Phi,\tau_i)$ of cosmological solutions. The supercritical regions satisfy $\abs c_\bot/c_\Phi\abs>\gamma_c$ and $\nF-\nB<0$. The subcritical region, $\abs c_\bot/c_\Phi\abs<\gamma_c$, contains an ellipse $\tau_-<\tau_i<\tau_+$, the interior (exterior) of it corresponding to models satisfying $\nF-\nB>0$ ($<0$). The trajectories $\tau(t)$ for increasing cosmic time $t$ are represented by dashed lines, for $c_\Phi>0$.}
\label{CI}
\end{figure}
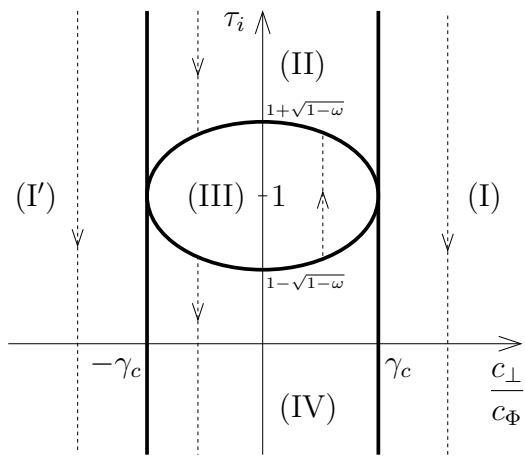

We have considered the low energy effective action of heterotic no-scale models compactified on tori down to $d$ dimensions. At 1-loop, the effective potential backreacts on the classical background, which is therefore time-dependent. Interested in homogeneous and isotropic cosmological evolutions, we have restricted our analysis to the dynamics of the scale factor $a(t)$, the no-scale modulus $\Phi(t)$ and a free scalar $\phi_\bot(t)$, which is a combination of the dilaton and the volume involved in the stringy Scherk-Schwarz supersymmetry breaking \cite{SSstring, Kounnas-Rostand}. The space of solutions can be parameterized by $(c_\bot/c_\Phi,\tau_i)$, where $c_\bot$, $c_\Phi$ are integration constants and $\tau_i$ is the initial value of $\mbox{$\tau(t)={2A\over dc_\Phi}\, \dot a \, a^{d-2}$}$ (see Eq.~(\ref{tau})). Fig.~\ref{CI} 
shows the partition of the $\R^2$-plane of cosmological solutions: The interior of the ellipse is realized by models where $\nF-\nB>0$, and yields trajectories $\tau(t)$ which follow dashed vertical lines,  from bottom to up when $c_\Phi>0$. Similarly, the exterior of the ellipse corresponds to models having $\nF-\nB<0$, with $\tau(t)$ following vertical lines from up to bottom when $c_\Phi>0$. 

The  trajectories in the supercritical regions (I) and (I$'$), which  have no classical counterparts, are characterized by a bounded scale factor. The evolution starts and ends with a Big Bang and a Big Crunch, where the universe is dominated by the kinetic energy and quantum potential of the no-scale modulus. Translated in terms of a perfect fluid of energy density $\rho$ and pressure $P$, we have
\be
\rho\sim {1\over 2}\, \dot\Phi^2+\kappa^2\Vone\, , \quad P\sim {1\over 2}\, \dot\Phi^2-\kappa^2\Vone\, , \quad P\sim \left({2A\over d-1}+1\right)\!\rho\, .
\ee
However, the above regime $\tau\to \pm \infty$ of low scale factor can only be trusted until higher order corrections in $\gs$ and $\alpha'$ become important.

On the contrary, the  solutions in the subcritical regions (II) and (III) are attracted to the quantum no-scale regime $\tau\to \tau_+$, which restores the no-scale structure \cite{noscale} as the universe expands, and is easily (if not always, see Eq. (\ref{1})) perturbative in $\gs$.  As a result, the evolution is asymptotically dominated by the classical kinetic energies of $\Phi$ and $\phi_\bot$. The endless expansion and flatness of the universe are compatible with quantum corrections, which justifies that rigid Minkowski spacetime may be postulated in quantum field theory. Moreover, the evolutions in the subcritical regions (III) and (IV)  admit the second quantum no-scale regime $\tau\to \tau_-$, which is realized as the scale factor tends (formally) to 0. In total, the trajectories in region (III) connect two regimes where 
\be
P\sim \rho \sim {1\over 2}\, \dot\Phi^2+{1\over 2}\, \dot\phi_\bot^2\, , 
\ee
with possibly an intermediate period of accelerated cosmology, however too short to account for inflation. In regions (II) and (IV), the state equation of the fluid evolves between $P\sim \rho$ and $\mbox{$P\sim \left({2A\over d-1}+1\right)\!\rho$}$.

On the one hand, the drop in $M(t)$, which takes place in the quantum no-scale regime $\mbox{$\tau\to \tau_+$}$ and is independent of the sign of the potential, forbids the existence of any cosmological constant \ie fluid satisfying $P\sim -\rho$. On the other hand, neglecting the time-evolution (which makes sense at a cosmological scale) of the scale factor and scalar fields to end up with a theory in rigid Minkowski spacetime valid today, the energy density $M^d$ is effectively constant but not coupled  to gravity. Thus, from either of these points of view, the words ``cosmological'' and ``constant'' exclude somehow each other.  Note that this is also the case in other frameworks. For instance, compactifying a string theory on a compact hyperbolic space, flat FLRW solutions can be found, where the volume of the internal space is time-dependent and plays formally the role of $M(t)$ in the present work. Its associated canonical field, which is similar to $\Phi$, admits an exponential and positive potential, however arising at tree level \cite{solV, DS, accel}. This setup can be realized by considering S-brane backgrounds or non-trivial fluxes.  In all these cases, it is important to study the constraints arising from variations  of couplings and masses at cosmological time scales, as well as present violations of the equivalence principle \cite{Damour:2002nv}.

The simple model we have analyzed in details in the present work can be upgraded in various ways. First of all, the full dependence of the 1-loop effective potential on the internal metric, internal antisymmetric tensor and  Wilson lines can be computed \cite{SNSM}. New effects then occur, due to the non-trivial metric of the moduli space and the existence of enhanced symmetry points \cite{CFP2}. Another direction of study consists in switching on finite temperature~$T$ \cite{Kounnas-Rostand, critical, attra, attraM/T, attT}. To see that the qualitative behaviour of the evolutions may be modified, let us assume $T\ll M$ and an expanding universe in quantum no-scale regime $\tau\to \tau_+$. In this case, we have \cite{attra}
\be
M^d={M^d_+\over a^{2(d-1)+K_+}} \, , \qquad T^d\sim {\#\over a^d}\, , 
\ee
so that $M/T$ decreases and the screening of thermal effects by quantum corrections eventually stops. In fact, new attractor mechanisms exist \cite{attra, attraM/T}. For instance, when $\nF-\nB>0$, quantum and thermodynamic corrections balance so that the free energy, which is nothing but the effective potential at finite temperature, yields a stabilization of $M(t)/T(t)$. At late times, the evolutions satisfy 
\be
{1\over a(t)}\sim \#\,  M(t) \sim \#\, T(t)\sim  \#\, e^{2\alpha^2\phi(t)}\sim {\#\over t^{2\over d}}\, ,
\ee
and are said ``radiation-like''. This is justified since the {\em total} energy density and pressure satisfy $\rho_{\rm tot}\sim (d-1)P_{\rm tot}$, where $\rho_{\rm tot}, P_{\rm tot}$ take into account the thermal energy density and pressure derived from the free energy, as well as the kinetic energy of the no-scale modulus~$\Phi$ \cite{critical, attra, attraM/T}. 


\section*{Acknowledgement}
 
We are grateful to  Steve Abel, Carlo Angelantonj, Keith Dienes, Emilian Dudas, Sergio Ferrara, Lucien Heurtier, Alexandros Kahagias and Costas Kounnas  for fruitful discussions. 
The work of H.P. is partially supported by the Royal Society International Cost Share Award. H.P. would like to thank the C.E.R.N. Theoretical Physics Department, the Simons Center for Geometry and Physics, and the IPPP in Durham University for hospitality.




\begin{thebibliography}{99}


\bibitem{Barbieri:1982eh}
  R.~Barbieri, S.~Ferrara and C.~A.~Savoy,
  ``Gauge models with spontaneously broken local supersymmetry,''
  Phys.\ Lett.\  B {\bf 119} (1982) 343;\\
  E.~Cremmer, P.~Fayet and L.~Girardello,
  ``Gravity induced supersymmetry breaking and low-energy mass spectrum,''
  Phys.\ Lett.\  B {\bf 122} (1983) 41.
  
  
     \bibitem{noscale}
  E.~Cremmer, S.~Ferrara, C.~Kounnas and D.~V.~Nanopoulos,
  ``Naturally vanishing cosmological constant in $\N=1$ supergravity,''
  Phys.\ Lett.\  B {\bf 133} (1983) 61;\\
  J.~R.~Ellis, C.~Kounnas and D.~V.~Nanopoulos,
  ``Phenomenological $SU(1,1)$ supergravity,''
  Nucl.\ Phys.\  B {\bf 241} (1984) 406;\\
  J.~R.~Ellis, A.~B.~Lahanas, D.~V.~Nanopoulos and K.~Tamvakis,
 ``No-scale supersymmetric standard model,''
  Phys.\ Lett.\  B {\bf 134} (1984) 429;\\
  J.~R.~Ellis, C.~Kounnas and D.~V.~Nanopoulos,
  ``No scale supersymmetric GUTs,''
  Nucl.\ Phys.\  B {\bf 247} (1984) 373.

\bibitem{Weinberg:1988cp}
  S.~Weinberg,
  ``The cosmological constant problem,''
  Rev.\ Mod.\ Phys.\  {\bf 61} (1989) 1.
  
    \bibitem{SSstring}
R.~Rohm,
``Spontaneous supersymmetry breaking in supersymmetric string theories,''
Nucl.\ Phys.\ B {\bf 237} (1984) 553;\\
C.~Kounnas and M.~Porrati,
``Spontaneous supersymmetry breaking in string theory,''
Nucl.\ Phys.\ B {\bf 310} (1988) 355;\\
  S.~Ferrara, C.~Kounnas and M.~Porrati,
``Superstring solutions with spontaneously broken four-dimensional
  supersymmetry,''
  Nucl.\ Phys.\  B {\bf 304} (1988) 500; \\
S.~Ferrara, C.~Kounnas, M.~Porrati and F.~Zwirner,
``Superstrings with spontaneously broken supersymmetry and their effective theories,''
Nucl.\ Phys.\ B {\bf 318} (1989) 75.

\bibitem{Kounnas-Rostand}
  C.~Kounnas and B.~Rostand,
  ``Coordinate-dependent compactifications and discrete symmetries,''
  Nucl.\ Phys.\ B {\bf 341} (1990) 641.

  \bibitem{SS}
  J.~Scherk and J.~H.~Schwarz,
  ``Spontaneous breaking of supersymmetry through dimensional reduction,''
  Phys.\ Lett.\  B {\bf 82} (1979) 60. 

  
    \bibitem{L=0}
  S.~Kachru, J.~Kumar and E.~Silverstein,
  ``Vacuum energy cancellation in a non-supersymmetric string,''
  Phys.\ Rev.\ D {\bf 59} (1999) 106004
  [hep-th/9807076]; \\
  G.~Shiu and S.~H.~H.~Tye,
  ``Bose-Fermi degeneracy and duality in non-supersymmetric strings,''
  Nucl.\ Phys.\ B {\bf 542} (1999) 45
  [hep-th/9808095];\\
  Y.~Satoh, Y.~Sugawara and T.~Wada,
  ``Non-supersymmetric asymmetric orbifolds with vanishing cosmological constant,''
  JHEP {\bf 1602} (2016) 184
  [arXiv:1512.05155 [hep-th]];\\
  Y.~Sugawara and T.~Wada,
  ``More on non-supersymmetric asymmetric orbifolds with vanishing cosmological constant,''
  JHEP {\bf 1608} (2016) 028
  [arXiv:1605.07021 [hep-th]].
  
   %
 \bibitem{1-L=0}
  R.~Blumenhagen and L.~Gorlich,
  ``Orientifolds of non-supersymmetric asymmetric orbifolds,''
  Nucl.\ Phys.\ B {\bf 551} (1999) 601
  [hep-th/9812158];\\
%
  C.~Angelantonj, I.~Antoniadis and K.~Forger,
  ``Non-supersymmetric type I strings with zero vacuum energy,''
  Nucl.\ Phys.\ B {\bf 555} (1999) 116
  [hep-th/9904092].
  %

\bibitem{GrootNibbelink:2017luf}
  S.~Groot Nibbelink, O.~Loukas, A.~Mütter, E.~Parr and P.~K.~S.~Vaudrevange,
  ``Tension between a vanishing cosmological constant and non-supersymmetric heterotic orbifolds,''
  arXiv:1710.09237 [hep-th].

  
 
\bibitem{Itoyama:1986ei}
  H.~Itoyama and T.~R.~Taylor,
  ``Supersymmetry restoration in the compactified $O(16) \times O(16)'$ heterotic string theory,''
  Phys.\ Lett.\ B {\bf 186} (1987) 129;\\
  S.~Abel, K.~R.~Dienes and E.~Mavroudi,
  ``Towards a non-supersymmetric string phenomenology,''
  Phys.\ Rev.\ D {\bf 91} (2015) 126014
  [arXiv:1502.03087 [hep-th]].
  
    \bibitem{SNSM}
  C.~Kounnas and H.~Partouche,
  ``Super no-scale models in string theory,''
  Nucl.\ Phys.\ B {\bf 913} (2016) 593
  [arXiv:1607.01767 [hep-th]];\\
  C.~Kounnas and H.~Partouche,
  ``$\mathcal N=2 \to 0$ super no-scale models and moduli quantum stability,''
  Nucl.\ Phys.\ B {\bf 919} (2017) 41
  [arXiv:1701.00545 [hep-th]].
  
  \bibitem{FR}
  I.~Florakis and J.~Rizos,
  ``Chiral heterotic strings with positive cosmological constant,''
  Nucl.\ Phys.\ B {\bf 913} (2016) 495
  [arXiv:1608.04582 [hep-th]].

  \bibitem{HarveyL=0}
  J.~A.~Harvey,
  ``String duality and non-supersymmetric strings,''
  Phys.\ Rev.\ D {\bf 59} (1999) 026002
  [hep-th/9807213].
  
   \bibitem{L2}
  K.~Aoki, E.~D'Hoker and D.~H.~Phong,
  ``Two-loop superstrings on orbifold compactifications,''
  Nucl.\ Phys.\ B {\bf 688} (2004) 3
  [hep-th/0312181];\\
  R.~Iengo and C.~J.~Zhu,
  ``Evidence for nonvanishing cosmological constant in nonSUSY superstring models,''
  JHEP {\bf 0004} (2000) 028
  [hep-th/9912074];\\
  S.~Abel and R.~J.~Stewart,
  ``On exponential suppression of the cosmological constant in non-SUSY strings at two loops and beyond,''
  Phys.\ Rev.\ D {\bf 96} (2017) 106013
  [arXiv:1701.06629 [hep-th]].
 
\bibitem{Damour:2002nv}
  T.~Damour, F.~Piazza and G.~Veneziano,
  ``Violations of the equivalence principle in a dilaton runaway scenario,''
  Phys.\ Rev.\ D {\bf 66} (2002) 046007
  [hep-th/0205111];\\
  T.~Damour, F.~Piazza and G.~Veneziano,
  ``Runaway dilaton and equivalence principle violations,''
  Phys.\ Rev.\ Lett.\  {\bf 89} (2002) 081601
  [gr-qc/0204094];\\
  T.~Damour and J.~F.~Donoghue,
  ``Equivalence principle violations and couplings of a light dilaton,''
  Phys.\ Rev.\ D {\bf 82} (2010) 084033
  [arXiv:1007.2792 [gr-qc]];\\
  T.~Damour and J.~F.~Donoghue,
  ``Phenomenology of the equivalence principle with light scalars,''
  Class.\ Quant.\ Grav.\  {\bf 27} (2010) 202001
  [arXiv:1007.2790 [gr-qc]].
  
  
  
\bibitem{solV}
  J.~J.~Halliwell,
  ``Scalar fields in cosmology with an exponential potential,''
  Phys.\ Lett.\ B {\bf 185} (1987) 341;\\
  A.~B.~Burd and J.~D.~Barrow,
  ``Inflationary models with exponential potentials,''
  Nucl.\ Phys.\ B {\bf 308} (1988) 929,
   Erratum: [Nucl.\ Phys.\ B {\bf 324} (1989) 276];\\
  L.~P.~Chimento,
  ``General solution to two-scalar field cosmologies with exponential potentials,''
  Class.\ Quant.\ Grav.\  {\bf 15} (1998) 965;\\
  I.~P.~C.~Heard and D.~Wands,
  ``Cosmology with positive and negative exponential potentials,''
  Class.\ Quant.\ Grav.\  {\bf 19} (2002) 5435
  [gr-qc/0206085];\\
  Z.~K.~Guo, Y.~S.~Piao and Y.~Z.~Zhang,
  ``Cosmological scaling solutions and multiple exponential potentials,''
  Phys.\ Lett.\ B {\bf 568} (2003) 1
  [hep-th/0304048];\\
  P.~K.~Townsend,
  ``Cosmic acceleration and M-theory,''
  hep-th/0308149;\\
  I.~P.~Neupane,
  ``Accelerating cosmologies from exponential potentials,''
  Class.\ Quant.\ Grav.\  {\bf 21} (2004) 4383
  [hep-th/0311071];\\
  P.~Vieira,
  ``Late-time cosmic dynamics from M-theory,''
  Class.\ Quant.\ Grav.\  {\bf 21} (2004) 2421
  [hep-th/0311173];\\
  E.~Bergshoeff, A.~Collinucci, U.~Gran, M.~Nielsen and D.~Roest,
  ``Transient quintessence from group manifold reductions or how all roads lead to Rome,''
  Class.\ Quant.\ Grav.\  {\bf 21} (2004) 1947
  [hep-th/0312102];\\
  J.~G.~Russo,
  ``Exact solution of scalar tensor cosmology with exponential potentials and transient acceleration,''
  Phys.\ Lett.\ B {\bf 600} (2004) 185
  [hep-th/0403010];\\
  L.~Jarv, T.~Mohaupt and F.~Saueressig,
  ``Quintessence cosmologies with a double exponential potential,''
  JCAP {\bf 0408} (2004) 016
  [hep-th/0403063];\\
  P.~K.~Townsend and M.~N.~R.~Wohlfarth,
  ``Cosmology as geodesic motion,''
  Class.\ Quant.\ Grav.\  {\bf 21} (2004) 5375
  [hep-th/0404241];\\
  A.~Collinucci, M.~Nielsen and T.~Van Riet,
  ``Scalar cosmology with multi-exponential potentials,''
  Class.\ Quant.\ Grav.\  {\bf 22} (2005) 1269
  [hep-th/0407047].
 
  
  
\bibitem{DS}
  E.~Dudas, N.~Kitazawa and A.~Sagnotti,
  ``On climbing scalars in string theory,''
  Phys.\ Lett.\ B {\bf 694} (2011) 80
  [arXiv:1009.0874 [hep-th]].
  
  \bibitem{accel}  
  P.~K.~Townsend and M.~N.~R.~Wohlfarth,
  ``Accelerating cosmologies from compactification,''
  Phys.\ Rev.\ Lett.\  {\bf 91} (2003) 061302
  [hep-th/0303097];\\
  N.~Ohta,
  ``Accelerating cosmologies from S-branes,''
  Phys.\ Rev.\ Lett.\  {\bf 91} (2003) 061303
  [hep-th/0303238];\\
  S.~Roy,
  ``Accelerating cosmologies from M/string theory compactifications,''
  Phys.\ Lett.\ B {\bf 567} (2003) 322
  [hep-th/0304084];\\
  M.~N.~R.~Wohlfarth,
  ``Accelerating cosmologies and a phase transition in M-theory,''
  Phys.\ Lett.\ B {\bf 563} (2003) 1
  [hep-th/0304089];\\
  R.~Emparan and J.~Garriga,
  ``A Note on accelerating cosmologies from compactifications and S-branes,''
  JHEP {\bf 0305} (2003) 028
  [hep-th/0304124];\\
  N.~Ohta,
  ``A Study of accelerating cosmologies from superstring/M theories,''
  Prog.\ Theor.\ Phys.\  {\bf 110} (2003) 269
  [hep-th/0304172];\\
  C.~M.~Chen, P.~M.~Ho, I.~P.~Neupane and J.~E.~Wang,
  ``A Note on acceleration from product space compactification,''
  JHEP {\bf 0307} (2003) 017
  [hep-th/0304177];\\
  C.~M.~Chen, P.~M.~Ho, I.~P.~Neupane, N.~Ohta and J.~E.~Wang,
  ``Hyperbolic space cosmologies,''
  JHEP {\bf 0310} (2003) 058
  [hep-th/0306291];\\
  M.~N.~R.~Wohlfarth,
  ``Inflationary cosmologies from compactification?,''
  Phys.\ Rev.\ D {\bf 69} (2004) 066002
  [hep-th/0307179].
 

\bibitem{attra}
  F.~Bourliot, C.~Kounnas and H.~Partouche,
  ``Attraction to a radiation-like era in early superstring cosmologies,''
  Nucl.\ Phys.\ B {\bf 816} (2009) 227
  [arXiv:0902.1892 [hep-th]].
  
  
\bibitem{Kiritsis:1994ta}
  E.~Kiritsis and C.~Kounnas,
  ``Infrared regularization of superstring theory and the one-loop calculation of coupling constants,''
  Nucl.\ Phys.\ B {\bf 442} (1995) 472
  [hep-th/9501020].
  
\bibitem{Faraggi:2014eoa}
  A.~E.~Faraggi, C.~Kounnas and H.~Partouche,
  ``Large volume susy breaking with a solution to the decompactification problem,''
  Nucl.\ Phys.\ B {\bf 899} (2015) 328
  [arXiv:1410.6147 [hep-th]].
  

\bibitem{CFP2}
  T.~Coudarchet and H.~Partouche,
  ``Quantum no-scale regimes and moduli dynamics,'' submitted for publication,
  arXiv:1804.00466 [hep-th].

  \bibitem{critical}
  T.~Catelin-Jullien, C.~Kounnas, H.~Partouche and N.~Toumbas,
 ``Thermal/quantum effects and induced superstring cosmologies,''
  Nucl.\ Phys.\ B {\bf 797} (2008) 137
  [arXiv:0710.3895 [hep-th]];\\
  T.~Catelin-Jullien, C.~Kounnas, H.~Partouche and N.~Toumbas,
  ``Induced superstring cosmologies and moduli stabilization,''
  Nucl.\ Phys.\ B {\bf 820} (2009) 290
  [arXiv:0901.0259 [hep-th]].
  
   \bibitem{attraM/T}
   F.~Bourliot, J.~Estes, C.~Kounnas and H.~Partouche,
  ``Cosmological phases of the string thermal effective potential,''
  Nucl.\ Phys.\ B {\bf 830} (2010) 330
  [arXiv:0908.1881 [hep-th]];\\
  J.~Estes, C.~Kounnas and H.~Partouche,
  ``Superstring cosmology for $\N_4 = 1 \to 0$ superstring vacua,''
  Fortsch.\ Phys.\  {\bf 59} (2011) 861
  [arXiv:1003.0471 [hep-th]].


   \bibitem{attT}
  J.~Estes, L.~Liu and H.~Partouche,
  ``Massless D-strings and moduli stabilization in type~I cosmology,''
  JHEP {\bf 1106} (2011) 060
  [arXiv:1102.5001 [hep-th]];\\
   L.~Liu and H.~Partouche,
  ``Moduli stabilization in type II Calabi-Yau compactifications at finite temperature,''
  JHEP {\bf 1211} (2012) 079
  [arXiv:1111.7307 [hep-th]].
  
  
  \end{thebibliography}
\end{document}